\documentclass[letterpaper,twoside]{article}
\usepackage[preprint]{aistats2027}
\usepackage{microtype}
\usepackage{amssymb}
\usepackage{amsthm}
\newtheorem{theorem}{Theorem}
\newtheorem{theorempart}{Theorem}[theorem]
\newtheorem{proposition}[theorem]{Proposition}
\newtheorem{corollary}[theorem]{Corollary}
\usepackage{graphicx}
\usepackage{enumitem}
\usepackage[round]{natbib}

\usepackage[hidelinks]{hyperref}
\usepackage{fontawesome5}
\newcommand{\mcmcCodeLink}{\\[0.6em]{\footnotesize
  \href{https://github.com/BaroqueObama/mcmc-impossibility-code}{\faGithub~Code}}}

\hypersetup{pdftitle={MCMC Can See Can't See},
pdfauthor={Daniel Matsui Smola}}

\begin{document}

\twocolumn[
\aistatstitle{MCMC Can See Can't See}
\vspace{-1.8em}
\aistatsauthor{Daniel Matsui Smola}
\aistatsaddress{Department of Computer Science\\
University of Washington, Seattle, WA\\
\href{mailto:daniel@matsuismola.com}{\texttt{daniel@matsuismola.com}}\mcmcCodeLink}
]

\begin{abstract}
It is folklore that passing ``convergence'' diagnostics does not guarantee that a Markov chain has converged. We prove this as an impossibility for any finite checklist of diagnostics. Running more chains, longer chains, and smarter diagnostics does not avoid this impossibility in certifying convergence. Under regularity assumptions, we determine how the worst-case undetected non-convergence decays as more diagnostics are added, and how to choose diagnostics optimally. Folklore suggests that the sampler's slowest mixing modes are the ``dangerous'' ones that must be monitored; we prove that the slow modes yield nearly optimal diagnostics as sampling proceeds. Given a source class, we also give an anytime-valid bound on the remaining discrepancy and show how undetected bias caps the effective sample size that a finite checklist can certify.

\end{abstract}

\section{Introduction}
\label{sec:introduction}

The typical Markov chain Monte Carlo (MCMC) workflow consists of running several chains until a checklist of ``convergence'' criteria are satisfied. A practitioner might require an $\hat R$ statistic less than 1.01 and an effective sample size (ESS) above 200 before using the output to estimate posterior expectations. Yet it has long been folklore that these checks do not guarantee convergence. \citet{cowles1996markov} document empirical failures of convergence diagnostics even in simple examples.

\begin{figure}[t]
\centering
\includegraphics[width=\columnwidth]{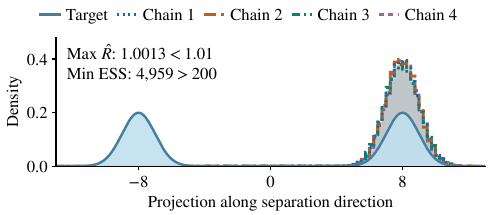}\par
\vspace{-4pt}
\setlength{\abovecaptionskip}{3pt}
\caption{Four MALA chains sampling an equally weighted Gaussian mixture. After 20,000 steps each, both diagnostics pass, but all chains miss the left mode.}
\label{fig:trapped-mixture}
\vspace{-11pt}
\end{figure}

Consider a target consisting of two well-separated Gaussian components with equal probability (Figure~\ref{fig:trapped-mixture}). Four chains using the Metropolis-adjusted Langevin algorithm (MALA) start in the same component and remain trapped there, missing half of the target distribution. Yet, across 30 repetitions, the largest $\hat R$ recorded is $1.0023 < 1.01$ and the smallest coordinate ESS is $4858>200$, satisfying both monitors. A practitioner may unknowingly consider these chains converged.

Despite these limitations, recent studies in \emph{Nature} and \emph{Science} still use a small set of monitors to assess convergence or select which chains to retain \citep{zanno2025nanotyrannus,bozzi2026treponema,livingston2026progenitor,ferrari2026landscapes}. Such checks can pass even when there is substantial error in posterior estimates. \emph{Would monitoring more statistics resolve this issue?}

We prove that no finite collection of monitors can certify convergence. The proof utilizes mean embeddings in a reproducing-kernel Hilbert space (RKHS) to express the chain's non-convergence and diagnostic monitors as vectors. Each diagnostic measures a dot product with the non-convergence vector; a value of zero means that it detects no discrepancy. All diagnostics can therefore pass while the chain differs from its target in directions orthogonal to the subspace the diagnostics span.

This yields the following contributions:
\begin{enumerate}[topsep=0pt,partopsep=0pt]
\item \textbf{Certification Impossibility.} No finite diagnostic checklist can certify convergence. We characterize the worst-case residual after $m$ optimally chosen diagnostics through a Kolmogorov $m$-width. The covariance operator's leading eigendirections attain this width (Section~3).

\item \textbf{Optimal Diagnostics.} We characterize the optimal monitors for the discrepancy remaining as you sample and show when monitors constructed from slow mixing modes approach this optimum. We also bound the loss from using approximate monitor directions (Section~4).

\item \textbf{No Escape.} Running more chains, sampling longer, and monitoring smarter diagnostics does not remove the certification impossibility (Section~5).

\item \textbf{Anytime Valid Bound.} Given a source class, we derive an anytime-valid upper bound on the remaining discrepancy (Section~6).

\item \textbf{Reliability Dual.} We show how undetected bias in posterior expectation estimates limits the effective sample size that a finite checklist of diagnostics can certify (Section~7).

\end{enumerate}

\section{Background}
\label{sec:background}

\subsection{Convergence Criteria as Expectations}
\label{sec:expectation-criteria}

Many convergence criteria compare two expectations, $\mathbb E_P[c(X)]-\mathbb E_Q[c(X)]=0$, for a test function $c$ on the chain's state space $\mathcal X$. The distributions $P$ and $Q$ depend on the monitor: Geweke compares an early segment of a chain with a late one, with $c$ a monitored coordinate \citep{geweke1992evaluating}; $\hat R$ combines between-chain and within-chain variation \citep{vehtari2021rank}. For fixed $c$, the map $P\mapsto\mathbb E_P[c(X)]$ is linear in the distribution $P$, even when $c$ is nonlinear in $x$. We will use kernels to exploit this linearity. Nonlinear combinations of these expectations are handled in Section~\ref{sec:nonlinear}.

\subsection{Kernels and Mean Embeddings}
\label{sec:kernels-embeddings}

Fix a positive-definite kernel $k:\mathcal X\times\mathcal X\to\mathbb R$ with RKHS $\mathcal H$ and feature map $\phi:\mathcal X\to\mathcal H$. The reproducing property gives $f(x)=\langle f,\phi(x)\rangle_{\mathcal H}$ for $f\in\mathcal H$, and $k(x,x')=\langle\phi(x),\phi(x')\rangle_{\mathcal H}$ \citep{scholkopf2002learning}. On a compact $\mathcal X$, $k$ is \emph{universal} if $\mathcal H$ is dense in $C(\mathcal X)$, so every continuous function can be uniformly approximated by RKHS functions \citep{sriperumbudur2011universality}.

For a distribution $P$ on $\mathcal X$ with $\mathbb E_P[\sqrt{k(X,X)}]<\infty$, the \emph{mean embedding} is $\mu_P:=\mathbb E_P[\phi(X)]\in\mathcal H$. The reproducing property yields $\mathbb E_P[f(X)]=\langle f,\mu_P\rangle_{\mathcal H}$ \citep{gretton2012kernel}. The kernel $k$ is \emph{characteristic} when $\mu_P=\mu_Q$ holds if and only if $P=Q$ \citep{sriperumbudur2010hilbert}; universality implies characteristic \citep{sriperumbudur2011universality}. The \emph{maximum mean discrepancy} (MMD) between $P$ and $Q$ is
\begin{equation}
\begin{aligned}
\operatorname{MMD}(P,Q)
&=\sup_{\|f\|_{\mathcal H}\le1}\bigl(\mathbb E_P[f]-\mathbb E_Q[f]\bigr)\\
&=\|\mu_P-\mu_Q\|_{\mathcal H},
\end{aligned}
\label{eq:mmd}
\end{equation}
where the \emph{witness function} $f^\star=(\mu_P-\mu_Q)/\|\mu_P-\mu_Q\|_{\mathcal H}$ attains the supremum for $\mu_P\ne\mu_Q$ \citep{gretton2012kernel}. For a characteristic kernel, $\operatorname{MMD}(P,Q)=0$ if and only if $P=Q$.

\subsection{Diagnostics as Directions}
\label{sec:diagnostic-directions}

For a criterion represented by $c(x)=\langle v,\phi(x)\rangle_{\mathcal H}$,
\begin{equation}
\mathbb E_P[c(X)]-\mathbb E_Q[c(X)]=\langle v,\mu_P-\mu_Q\rangle_{\mathcal H}.
\label{eq:criterion-dot-product}
\end{equation}
For a comparison between the chain's law $\pi_t$ and its target $\pi$, let $\delta_t:=\mu_{\pi_t}-\mu_\pi$ be the \emph{non-convergence embedding}. Then, passing a criterion means $\langle v,\delta_t\rangle_{\mathcal H}=0$. By construction, $\operatorname{MMD}(\pi_t,\pi)=\|\delta_t\|_{\mathcal H}$. Comparisons between chains and between segments of a chain are discussed in Section~5.1.

\subsection{Related Work}
\label{sec:related-work}

\citet{asmussen1992stationarity} rule out stationarity detection from finite simulation output uniformly over countably infinite-state ergodic chains. We fix the target and study what finite checklists miss even with exact expectation readings.

\citet{manole2023martingale} give anytime-valid bounds on MMD between two distributions from independent samples. Other methods test agreement with a target using Stein discrepancy \citep{martinez2025sequential} or expectations estimated from MCMC output \citep{gyori2016hypothesis}. Section~6 also bounds the component a finite checklist leaves unmeasured.

Kernel methods estimate slow mixing modes \citep{klus2020eigendecompositions,tian2020kvad}. For the canonical diffusion, \citet{tolkovsky2026mode} show that leading autocovariance subspaces approach the span of embedded slow modes. Section~4 connects this alignment to nearly optimal diagnostics for the worst undetected discrepancy.

\section{Certification Impossibility}
\label{sec:certification}

\subsection{Qualitative Theorem}
\label{sec:qualitative}

We return to the question raised in the introduction: \emph{can a finite checklist certify convergence?} We prove that it cannot, even when every criterion is satisfied exactly.

\begin{theorem}[Certification impossibility]
\label{thm:certification}
Let $\pi$ have a density on $\mathbb R^d$, and let $k$ be a bounded characteristic kernel with RKHS $\mathcal H$. Let the checklist $v_1,\ldots,v_m\in\mathcal H$ consist of $m$ diagnostic directions, and set $V:=\operatorname{span}\{v_1,\ldots,v_m\}$. Satisfying every criterion means $\langle v_i,\delta_t\rangle_{\mathcal H}=0$ for all $i=1,\ldots,m$ (equivalently $\delta_t\perp V$).

Then, satisfying these criteria cannot rule out a non-converged chain: for any such finite checklist, an ergodic chain targeting $\pi$ can satisfy, from a suitable initialization and at every finite time $t$,
\begin{equation}
\delta_t\perp V\quad\text{and}\quad\delta_t\ne0\quad\text{(hence $\pi_t\ne\pi$)},
\label{eq:passing-without-convergence}
\end{equation}
where an explicit monitor to detect this full discrepancy is the MMD witness $v_{m+1}:=\delta_t/\|\delta_t\|_{\mathcal H}$, satisfying
\begin{equation}
\langle v_{m+1},\delta_t\rangle_{\mathcal H}
=\|\delta_t\|_{\mathcal H}=\operatorname{MMD}(\pi_t,\pi)>0.
\label{eq:unmonitored-witness}
\end{equation}
\end{theorem}

\begin{proof}[Proof sketch]
Moving $\mu_{\pi_t}$ away from $\mu_\pi$ in a direction orthogonal to $V$ leaves every monitored projection unchanged. Thus both a converged $\pi_t$ ($\delta_t=0$) and a non-converged $\pi_t$ (where $\delta_t\ne0$ and $\delta_t\perp V$) give the same zero readings for the diagnostics. The checklist therefore cannot distinguish a converged chain from a non-converged one. Appendix~\ref{app:qualitative} constructs an ergodic chain realizing the latter case at every finite time. Normalizing $\delta_t$ gives the MMD witness.
\end{proof}
\paragraph{Implications.}
Running all chains until every criterion is satisfied does not guarantee convergence. Averaging an unmonitored function $f(X)$ over the samples can still give a biased estimate of $\mathbb E_\pi[f(X)]$.

\subsection{Quantitative Theorem}
\label{sec:quantitative}

Satisfying more monitors should nonetheless buy us something. \emph{How large can the worst-case remaining discrepancy be if we satisfy the best $m$ monitors?}

Although $\mathcal H$ is infinite-dimensional, variation under the target $\pi$ may concentrate in relatively few directions. The target distribution's covariance operator ranks these directions by variance:
\begin{equation}
\Sigma:=\mathbb E_\pi\bigl[(\phi(X)-\mu_\pi)\otimes(\phi(X)-\mu_\pi)\bigr].
\label{eq:covariance}
\end{equation}
For bounded $k$, the eigendecomposition of $\Sigma$ gives eigenvalues $\lambda_1\ge\lambda_2\ge\cdots\ge0$ and orthonormal eigenvectors $e_1,e_2,\ldots$ on its positive spectrum. We first use a standard source condition \citep{caponnetto2007optimal}:

\noindent\textbf{(A1) Source condition.} $\delta_t=\Sigma^r u$ with $\|u\|_{\mathcal H}\le R$, for fixed $r,R>0$.

This limits how much discrepancy can lie in directions with small eigenvalues. These vectors form the ellipsoid $\mathcal B_r(R):=\{\Sigma^r u:\|u\|_{\mathcal H}\le R\}$, with semi-axes $R\lambda_j^r$. The bound below is exact over this ellipsoid; its worst case need not be realized by a probability law (Appendix~\ref{app:quantitative}). Write $P_V$ for orthogonal projection onto $V$, so that $P_{V^\perp}\delta_t$ is the remaining discrepancy.

\refstepcounter{theorem}
\begin{theorempart}[Optimal residual bound]
\label{thm:quantitative}
Under \textbf{(A1)},
\begin{equation}
\inf_{\dim V\le m}\ \sup_{\delta\in\mathcal B_r(R)}
\|P_{V^\perp}\delta\|_{\mathcal H}=R\lambda_{m+1}^r,
\label{eq:width}
\end{equation}
attained by the $m$ optimal monitors $e_1,\ldots,e_m$, with span $V_m^\star$. The value is unchanged if the supremum is restricted to $\delta\in V^\perp$, so this worst case remains even when all monitors are satisfied.
\end{theorempart}
\begin{proof}[Proof sketch]
Monitoring the first $m$ eigendirections removes the $m$ longest axes of the ellipsoid, leaving a residual of at most the next longest semi-axis, $R\lambda_{m+1}^r$. By pigeonhole, any $m$-dimensional subspace leaves a direction in the span of the first $m+1$ eigenvectors unmonitored. The ellipsoid extends at least $R\lambda_{m+1}^r$ along the unmonitored direction, giving the matching lower bound (Appendix~\ref{app:quantitative}).
\end{proof}

Thus, satisfying the best $m$ monitors restricts the worst-case residual to $R\lambda_{m+1}^r$. To determine how quickly this bound decays as we add monitors, we make a second standard assumption:

\noindent\textbf{(A2) Polynomial eigendecay.} $\lambda_j\asymp j^{-\alpha}$ for some $\alpha>1$.

\begin{theorempart}[Rate with additional monitors]
\label{thm:quantitative-rate}
Under \textbf{(A1)} and \textbf{(A2)}, the optimal residual in~\eqref{eq:width} satisfies
\[
R\lambda_{m+1}^r=\Theta(Rm^{-\alpha r}).
\]
\end{theorempart}

\paragraph{Implications.}
The optimal residual is the \emph{Kolmogorov $m$-width} of the source class \citep{pinkus1985nwidths}. As we add monitors, the residual decays at rate $m^{-\alpha r}$; larger $\alpha$ (faster spectral decay) or $r$ (greater smoothness) make it decay faster. Nevertheless, under \textbf{(A2)} the worst case remains nonzero for every finite $m$, even when all monitors are satisfied.

\paragraph{When does the source condition hold?}
Write $L^2(\pi)$ for the functions with finite second moment under $\pi$, with norm $\|h\|_{L^2(\pi)}=(\mathbb E_\pi[h(X)^2])^{1/2}$. If $h=d\pi_t/d\pi-1$ belongs to this space, then \textbf{(A1)} holds with $r=1/2$ and $R=\|h\|_{L^2(\pi)}$. An $L^p$ density error, $1<p<2$, gives the weaker exponent $r=1-1/p$ (Appendix~\ref{app:source-conditions}).

\subsection{Inexact Diagnostics}
\label{sec:inexact}

Practical diagnostics allow small deviations from their target values. Let $\eta$ bound the monitored deviations, $\|P_V\delta_t\|_{\mathcal H}\le\eta$. For orthonormal monitors, individual tolerances $|\langle v_i,\delta_t\rangle_{\mathcal H}|\le\varepsilon$ give $\eta=\sqrt m\,\varepsilon$. The orthogonal component satisfies
\begin{equation}
\begin{aligned}
\|P_{V^\perp}\delta_t\|_{\mathcal H}^2
&=\operatorname{MMD}^2(\pi_t,\pi)-\|P_V\delta_t\|_{\mathcal H}^2\\
&\ge\operatorname{MMD}^2(\pi_t,\pi)-\eta^2.
\end{aligned}
\label{eq:residual-decomposition}
\end{equation}
Thus, if $\operatorname{MMD}(\pi_t,\pi)>\eta$, a nonzero discrepancy remains in $V^\perp$ that the monitors do not detect. The certification impossibility persists (Appendix~\ref{app:inexact}).

\section{Optimal Diagnostics}
\label{sec:spectral-duality}

The optimal monitors in Theorem~\ref{thm:quantitative} depend on the target's covariance operator $\Sigma$, which is generally unknown. We do, however, know the sampler's transition rule. It is folklore that slow mixing modes are important to monitor because their bias persists longest. \emph{Can the sampler therefore guide us toward optimal diagnostics?}

\subsection{Optimal Monitors at Time \texorpdfstring{$t$}{t}}
\label{sec:dynamic-monitors}

Section~3 chooses monitors using a source condition alone. Here we use both a bound on the initial error and the sampler's transition rule to determine which discrepancies can remain at time $t$. \emph{Which $m$ monitors, when satisfied, give the smallest worst-case residual at time $t$?}

Fix a sampler with invariant law $\pi$ and retain the bounded characteristic kernel from Section~3. Let $\pi_0$ denote the sampler's distribution at time zero. We bound its initial relative density error:

\noindent\textbf{(A3) Bounded initialization error.} $\pi_0\ll\pi$, with $h_0:=d\pi_0/d\pi-1$ satisfying $\mathbb E_\pi[h_0(X)^2]\le R_{\mathrm{init}}^2$.

Here $h_0$ is the \emph{centered density ratio}, or the Radon--Nikodym derivative minus one. For continuous targets, fixed point initializations fall outside \textbf{(A3)} and require a separate argument (Appendix~\ref{app:source-conditions}).

Let $F_t(x)$ be the discrepancy embedding for a chain initialized at $x$. Define its covariance $\Sigma_t$ under $X\sim\pi$:
\begin{equation}
\begin{aligned}
F_t(x)&:=\mathbb E[\phi(X_t)\mid X_0=x]-\mu_\pi,\\
\Sigma_t&:=\mathbb E_\pi[F_t(X)\otimes F_t(X)].
\end{aligned}
\label{eq:dynamic-covariance}
\end{equation}
Averaging over the actual initial law gives
\[
\delta_t=\mathbb E_{\pi_0}[F_t(X)]=\mathbb E_\pi[h_0(X)F_t(X)].
\]
The mean-zero functions $h_0$ satisfying the norm bound in \textbf{(A3)} form a ball of radius $R_{\mathrm{init}}$ in $L^2(\pi)$. The equation above maps each initial error $h_0$ linearly to its time-$t$ discrepancy. This maps the ball to an ellipsoid $\mathcal B^{(t)}(R_{\mathrm{init}})$ containing every allowed $\delta_t$. Write $\lambda_{1,t}\ge\lambda_{2,t}\ge\cdots\ge0$ for the eigenvalues of $\Sigma_t$. As in Section~3, its eigenvectors give the ellipsoid's axes, with semi-axis lengths $R_{\mathrm{init}}\sqrt{\lambda_{j,t}}$.

At $t=0$, we recover $\Sigma_0=\Sigma$ and Section~3's source class $\mathcal B_{1/2}(R_{\mathrm{init}})$. Applying the sampler restricts this class to $\mathcal B^{(t)}(R_{\mathrm{init}})\subseteq\mathcal B_{1/2}(R_{\mathrm{init}})$. As initial errors decay at different rates, the ellipsoid can shrink unevenly and its longest axes can change direction. These axes identify where the greatest bias can still remain at time $t$. Theorem~\ref{thm:quantitative}'s ellipsoid argument now applies: monitoring the longest $m$ axes leaves a worst-case residual equal to the next longest semi-axis (Appendix~\ref{app:dynamic-width}).

\begin{proposition}[Optimal monitors at time $t$]
\label{thm:dynamic-width}
For the ellipsoid $\mathcal B^{(t)}(R_{\mathrm{init}})$, let $d_m(t)$ be the worst discrepancy that can remain undetected at time $t$ when the best $m$ monitors are satisfied:
\begin{equation}
\begin{aligned}
d_m(t)&:=\inf_{\dim V\le m}\ \sup_{\substack{\delta\in\mathcal B^{(t)}(R_{\mathrm{init}})\\\delta\perp V}}\|\delta\|_{\mathcal H}\\
&=R_{\mathrm{init}}\sqrt{\lambda_{m+1,t}}.
\end{aligned}
\label{eq:dynamic-width}
\end{equation}
The optimal monitor span $V_m^\star(t)$ consists of the leading $m$ eigenvectors of $\Sigma_t$.
\end{proposition}

\paragraph{Implications.}
Accounting for the sampler can therefore tighten Section~3's $r=1/2$ bound. For a reversible sampler, $\Sigma_t$ can be estimated from pairs of states along a stationary chain:
\[
\Sigma_t=\mathbb E_\pi\bigl[(\phi(X_0)-\mu_\pi)\otimes(\phi(X_{2t})-\mu_\pi)\bigr].
\]
The two copies of $F_t$ in the covariance correspond to two $t$-step runs from the same $X\sim\pi$, independent given $X$. Reversibility lets us reverse the first run and join it to the second, giving one stationary path with $t+t$ transitions (Appendix~\ref{app:dynamic-width}). Estimation is discussed in Section~\ref{sec:approximate-monitors}.

\subsection{Slow Modes Become Nearly Optimal}
\label{sec:optimal-mixing-modes}

\emph{Can the sampler's slow modes give a checklist as informative as the leading eigenvectors of $\Sigma_t$?} A \emph{mixing mode} is a function whose expectation bias is multiplied by the same factor at each transition. For the transition operator $(\mathcal P f)(x):=\mathbb E[f(X_1)\mid X_0=x]$, a mode $\psi_j$ is a mean-zero eigenfunction satisfying $\mathcal P\psi_j=\rho_j\psi_j$, where $|\rho_j|<1$. Thus
\[
\mathbb E_{\pi_t}[\psi_j(X)]=\rho_j^t\mathbb E_{\pi_0}[\psi_j(X)].
\]
Modes with $|\rho_j|$ close to one are \emph{slow modes}: their expectation bias decays slowly. Reversibility makes $\mathcal P$ self-adjoint, allowing an orthonormal choice of these eigenfunctions in $L^2(\pi)$ \citep[Lemma~3.9]{rudolf2012explicit}.

For a reversible sampler, the discrepancy associated with mode $\psi_j$ lies along the RKHS direction
\begin{equation}
w_j:=\mathbb E_\pi[(\phi(X)-\mu_\pi)\psi_j(X)].
\label{eq:slow-directions}
\end{equation}
The direction $w_j$ is fixed; time changes the amplitude of this component through $\rho_j^t$. We select the $m$ slowest transition modes, monitor $W_m:=\operatorname{span}\{w_1,\ldots,w_m\}$, and require a gap separating them from the remaining modes:

\noindent\textbf{(A4) Separation of slow modes.} For $0\le a<b<1$, the selected $m$ modes satisfy $|\rho_j|\ge b$, while every mean-zero $f$ orthogonal to them satisfies $\|\mathcal P f\|_{L^2(\pi)}\le a\|f\|_{L^2(\pi)}$.

The selected transition modes retain factors of magnitude at least $b^t$, while the remaining part decays at least as fast as $a^t$. For a finite ergodic reversible chain, \textbf{(A4)} holds at a strict gap $|\rho_m|>|\rho_{m+1}|$, where the $\rho_j$ are the eigenvalues of the transition operator $\mathcal P$ on mean-zero functions, ordered by magnitude.

\begin{theorem}[Slow-mode diagnostics]
\label{thm:slow-optimality}
For a reversible sampler satisfying \textbf{(A3)} and \textbf{(A4)}, for all sufficiently large $t$,
\begin{equation}
\sup_{\substack{\delta\in\mathcal B^{(t)}(R_{\mathrm{init}})\\\delta\perp W_m}}\|\delta\|_{\mathcal H}
\le(1+\varepsilon_t)d_m(t),\qquad \varepsilon_t\to0.
\label{eq:slow-optimality}
\end{equation}
\end{theorem}

\begin{proof}[Proof sketch]
By \textbf{(A4)}, omitted components shrink relative to the selected modes at rate $(a/b)^t$. The leading eigendirections of $\Sigma_t$ therefore approach $W_m$, whose worst-case residual is at most $1+o(1)$ times the optimum $d_m(t)$ in Proposition~\ref{thm:dynamic-width} (Appendix~\ref{app:slow-optimality}).
\end{proof}

\paragraph{Implications.}
The optimal space $V_m^\star(t)$ approaches the fixed span $W_m$ of embedded slow modes. Once these modes are estimated accurately, the same monitors can be nearly optimal at sufficiently late inspection times.

\subsection{Approximate Monitors}
\label{sec:approximate-monitors}

\emph{How does imperfect estimation of the monitor directions affect the residual bound?} Let $W$ be the chosen $m$-dimensional space and $V_m^\star$ the optimal span from Section~3. Their \emph{largest principal angle} satisfies $\cos\theta=\min_{v\in V_m^\star,\,\|v\|=1}\|P_Wv\|$ \citep{bjorck1973numerical}. The minimum selects the optimal direction least represented in $W$. An angle of zero means the spaces coincide; $90^\circ$ means an optimal direction is completely missed.

\begin{corollary}[Residual bound for approximate monitors]
\label{thm:approximate-monitors}
For $\delta_t\in\mathcal B_r(R)$ and $\theta<\pi/2$,
\begin{equation}
\delta_t\perp W\quad\Longrightarrow\quad
\|\delta_t\|_{\mathcal H}\le R\lambda_{m+1}^r\sec\theta.
\label{eq:approximate-monitors}
\end{equation}
For $\mathcal B^{(t)}(R_{\mathrm{init}})$, replace $R\lambda_{m+1}^r$ by $d_m(t)$ and measure the angle to $V_m^\star(t)$.
\end{corollary}

\paragraph{Implications.}
A fixed angle below $90^\circ$ preserves the width rate, so approximate monitors suffice with a constant penalty. An angle of $30^\circ$, for example, enlarges the bound by at most 15.5\%. This geometry also appears in recovery from subspace measurements \citep{binev2015data}; Appendix~\ref{app:approximate-monitors} gives the proof and tolerance bound.

\paragraph{Estimating the slow modes.}
Time-lagged independent component analysis (TICA) and kernel transfer-operator methods estimate $\psi_j$ from transitions \citep{perez2013identification,klus2020eigendecompositions}. Substituting the estimated mode into~\eqref{eq:slow-directions} and replacing its target expectation by a sample average estimates $w_j$. For a reversible sampler, one can also estimate $\Sigma_t$ from states separated by $2t$ transitions and take its leading eigenspace $V_m^\star(t)$ \citep{mollenhauer2022autocovariance}. Agreement between estimates cannot reveal a region both pilots missed (Appendix~\ref{app:mode-estimation}).

\section{No Escape}
\label{sec:no-escape}

\emph{Could more chains, smarter diagnostics, or more samples make a finite checklist sufficient?}

\subsection{More Chains}
\label{sec:more-chains}

Running several chains can reveal non-convergence when the runs disagree, as diagnostics such as $\hat R$ seek to detect \citep{vehtari2021rank}. We show how multiple chains can miss a shared discrepancy. Let $\delta_i$ be chain $i$'s discrepancy and $\bar\delta=\frac{1}{C}\sum_{i=1}^C\delta_i$ be the average discrepancy. Then
\begin{equation}
\frac1C\sum_{i=1}^C\|\delta_i\|^2
=\|\bar\delta\|^2+
\frac1C\sum_{i=1}^C\|\delta_i-\bar\delta\|^2.
\label{eq:chain-anova}
\end{equation}
The left-hand side is the average squared discrepancy from the target. The last term measures disagreement between chains; the first term, $\|\bar\delta\|^2$, measures their shared discrepancy. If all chains have the same law, the last term is zero: they agree with one another, but may still differ from the target. Such was the case in Figure~\ref{fig:trapped-mixture}, where every chain misses the same mode. Geweke's comparison of early and late chain segments has the same limitation: both segments can share a discrepancy from the target \citep{geweke1992evaluating}.

\subsection{Smarter Diagnostics}
\label{sec:nonlinear}

So far, our diagnostic readings have been linear in the distribution. \emph{Could nonlinear diagnostics certify convergence?} A convergent Taylor expansion in the distribution, called a \emph{von Mises expansion}, expresses nonlinear diagnostics through moment differences \citep{fernholz1983von}. Writing $\Delta_j:=\mathbb E_{\pi_t}[c_j(X)]-\mathbb E_\pi[c_j(X)]$, the expansion for each diagnostic $T_\ell$ takes the form
\[
\begin{aligned}
T_\ell(\pi_t)-T_\ell(\pi)
&=\sum_j a_{\ell j}\Delta_j\\
&\quad+\frac12\sum_{j,k}b_{\ell jk}\Delta_j\Delta_k+\cdots.
\end{aligned}
\]
Each $\Delta_j$ is again an expectation comparison, so we can treat it as a monitor. If finitely many moment functions appear across all the diagnostics, we are back to a finite checklist: matching those moments makes every term zero, even when the chain has not converged. Variance, for example, only requires matching the first two moments.

An expansion may involve infinitely many moments. If their closed span is not dense, an orthogonal complement remains; any nonzero discrepancy realizable there is still invisible to every term. (Think of Hilbert's hotel: monitoring every other eigendirection still leaves infinitely many unmonitored.) Thus the number of terms alone does not resolve the impossibility. Exact MMD uses the whole RKHS unit ball, whose span is all of $\mathcal H$, leaving no unmonitored orthogonal component. Appendix~\ref{app:nonlinear-moments} gives the expansion conditions and probability-law constructions.

\subsection{More Samples}
\label{sec:trajectories}

\emph{Would running chains longer resolve the problems?} More samples can reduce estimation error in the diagnostic readings. Even when these diagnostics are known without estimation error, the impossibility in Theorem~\ref{thm:certification} holds. Thus, running the chains until every criterion is satisfied cannot rule out a discrepancy orthogonal to the diagnostic span. Longer sampling can bring the chain closer to its target, but satisfying the checklist still does not establish that it has converged.

\section{Anytime Bounds}
\label{sec:anytime}

\emph{Can estimated diagnostic readings tell us when the discrepancy is below a chosen tolerance?} Sections~3 and~4 bound the unmonitored component. We combine these bounds with estimation uncertainty in the diagnostic readings and require the guarantee to hold at every inspection time, so that the readings can determine when to stop sampling the chains.

\subsection{A Bound Given the Source Class}
\label{sec:residual-bound}

Let $V_t$ be the diagnostic span used at time $t$. Let $M_t$ bound the monitored component $\|P_{V_t}\delta_t\|$, including observed deviations and estimation uncertainty, and let $U_t$ bound the unmonitored component $\|P_{V_t^\perp}\delta_t\|$.

Choose a total failure probability $\beta\in(0,1)$ for the monitoring procedure. It sets the confidence radii in $M_t$ and any estimated $U_t$.

\begin{theorem}[Anytime residual bound]
\label{thm:anytime}
Suppose both component bounds hold simultaneously at every inspection time with probability at least $1-\beta$. Then, on the same event, at every inspection time,
\begin{equation}
\operatorname{MMD}(\pi_t,\pi)
\le\sqrt{M_t^2+U_t^2}.
\label{eq:anytime-residual}
\end{equation}
\end{theorem}

\begin{proof}[Proof sketch]
Pythagoras gives $\|\delta_t\|^2=\|P_{V_t}\delta_t\|^2+\|P_{V_t^\perp}\delta_t\|^2$. Substitute the two bounds (Appendix~\ref{app:combined-bound}).
\end{proof}

\paragraph{Baseline: Section 3.}
If the source condition \textbf{(A1)} holds at every inspection time, use the optimal span $V_t=V_m^\star$ and $U_t=R\lambda_{m+1}^r$. This bounds the unmonitored discrepancy using the source condition alone, without information about the sampler's transition rule.

\paragraph{Refinement: Section 4.}
Under \textbf{(A3)}, use $V_t=V_m^\star(t)$ and $U_t=d_m(t)=R_{\mathrm{init}}\sqrt{\lambda_{m+1,t}}$. The sampler restricts which discrepancies can remain at time $t$. Assumption \textbf{(A3)} also implies \textbf{(A1)} with $r=1/2$ and $R=R_{\mathrm{init}}$, for which $d_m(t)\le R_{\mathrm{init}}\sqrt{\lambda_{m+1}}$. The unmonitored bound can therefore shrink as sampling continues, even with the same number of diagnostics. Whether the full bound is smaller also depends on $M_t$, which can change with the diagnostic span.

\paragraph{Implications.}
Choose a tolerance $\varepsilon>0$ and stop sampling at an inspection time when $\sqrt{M_t^2+U_t^2}\le\varepsilon$. With probability at least $1-\beta$, this rule never stops with $\operatorname{MMD}(\pi_t,\pi)>\varepsilon$. As the monitored deviations and their estimation uncertainty vanish, the bound approaches $U_t$. Section~4 can lower this floor as sampling continues, allowing smaller tolerances than Section~3's fixed bound.

\subsection{Using the Bound}

To obtain $M_t$, seek an estimate $\widehat\delta_t$ satisfying $\|\widehat\delta_t-\delta_t\|\le\eta_t$ simultaneously at every inspection time. Then $M_t=\|P_{V_t}\widehat\delta_t\|+\eta_t$. A projected-error bound $\|P_{V_t}(\widehat\delta_t-\delta_t)\|\le\eta_t$ also suffices.

The error bound $\eta_t$ must cover uncertainty in both chain and target expectations. For bounded kernels, Appendix~\ref{app:replicas} constructs it using independent chain replicas and a target reference with a known error bound. Independent target draws provide such a reference; a chain satisfying diagnostics alone does not.

For $U_t$, use Theorem~\ref{thm:quantitative} under \textbf{(A1)} or Proposition~\ref{thm:dynamic-width} under \textbf{(A3)}. In the latter case, a positive covariance estimate $\widehat\Sigma_t$ with operator error at most $\epsilon_{\Sigma,t}$ gives $U_t=R_{\mathrm{init}}\sqrt{\widehat\lambda_{m+1,t}+\epsilon_{\Sigma,t}}$ when the diagnostics span its leading $m$ eigendirections; $\widehat\lambda_{m+1,t}$ is the next eigenvalue. Appendix~\ref{app:covariance-estimation} gives covariance confidence bounds for bounded kernels using independent target draws. A single trajectory requires quantitative dependence assumptions \citep{mollenhauer2022autocovariance}. Source parameters still need independent justification, and all statistical bounds must share the failure budget $\beta$.

\paragraph{Can the source assumption be avoided?}
The full-norm error bound also gives $\operatorname{MMD}(\pi_t,\pi)\le\|\widehat\delta_t\|+\eta_t$, without a source radius or covariance estimate. Neither this direct bound nor~\eqref{eq:anytime-residual} is uniformly tighter. Our conditional bound can exploit a small $U_t$ justified by source or sampler information, and also works with only projected-error bounds. Additional residual measurements offer another route (Appendix~\ref{app:residual-energy}).

\section{Reliability Dual}
\label{sec:reliability}

Effective sample size (ESS) is the number of independent target draws that would give the same sampling variance for an estimate \citep{vehtari2021rank}. \emph{Does a large ESS guarantee an accurate posterior estimate?} ESS accounts for dependence between samples, and a larger value means lower sampling variance for the same quantity. It does not account for initialization bias. We ask how much accuracy a satisfied checklist can guarantee when this bias is included.

Discard the first $b$ states as \emph{burn-in} to reduce the effect of initialization, and retain the following $n$ states for estimation. Their average $\bar f_{b,n}:=\frac1n\sum_{s=0}^{n-1}f(X_{b+s})$ estimates $\mathbb E_\pi[f(X)]$. Its mean-squared error is
\[
\operatorname{MSE}(\bar f_{b,n})
=\operatorname{Var}(\bar f_{b,n})+\operatorname{Bias}(\bar f_{b,n})^2.
\]
For $f\in\mathcal H$, the bias is $\langle f,\bar\delta_{b,n}\rangle$, where $\bar\delta_{b,n}:=\frac1n\sum_{s=0}^{n-1}\delta_{b+s}$ is the average discrepancy over the retained times. Thus $\|\bar\delta_{b,n}\|$ is the largest absolute bias over unit-norm RKHS functions. The criterion for diagnostic $v_i$ is satisfied when $\langle v_i,\bar\delta_{b,n}\rangle=0$.

To determine the accuracy this checklist can guarantee, we need the worst bias it can leave in the sample average. Recall that $F_t(x)$ in Section~4 is the discrepancy embedding at time $t$ for a chain initialized at $x$. Average it over the retained times: $\bar F_{b,n}(x):=\frac1n\sum_{s=0}^{n-1}F_{b+s}(x)$. Define its covariance under $X\sim\pi$ by $\Sigma_{b,n}:=\mathbb E_\pi[\bar F_{b,n}(X)\otimes \bar F_{b,n}(X)]$, with eigenvalues $\lambda_{1,b,n}\ge\lambda_{2,b,n}\ge\cdots$. Since $\bar\delta_{b,n}=\mathbb E_\pi[h_0(X)\bar F_{b,n}(X)]$, \textbf{(A3)} again gives an ellipsoid, now describing the possible bias of retained averages.

\begin{corollary}[Precision after satisfying the diagnostics]
\label{cor:reliability}
For this ellipsoid, the worst-case bias remaining after the best $m$ diagnostics are satisfied is
\begin{equation}
d_{b,n,m}=R_{\mathrm{init}}\sqrt{\lambda_{m+1,b,n}}.
\label{eq:average-width}
\end{equation}
For the chosen kernel $k$, let $\sigma_k^2=\|\Sigma\|_{\mathrm{op}}>0$ be the largest target variance of a unit-norm RKHS function.

Assume each checklist can miss a bias of this size for some valid initialization under \textbf{(A3)}. The checklist then cannot guarantee mean-squared error below $d_{b,n,m}^2$ across all allowed initializations and unit-norm functions. Expressed as the accuracy of $N_{\mathrm{cert}}$ independent target draws, this gives
\begin{equation}
N_{\mathrm{cert}}\le\frac{\sigma_k^2}{d_{b,n,m}^2},
\qquad d_{b,n,m}>0.
\label{eq:certifiable-ess}
\end{equation}
\end{corollary}

\begin{proof}[Proof sketch]
Section~3's ellipsoid argument gives~\eqref{eq:average-width}. A hidden bias of size $d_{b,n,m}$ gives mean-squared error at least $d_{b,n,m}^2$ for its unit witness. In comparison, an average of $N$ independent target draws is unbiased and has variance $\operatorname{Var}_\pi(f(X))/N$. Its worst-case mean-squared error is therefore $\sigma_k^2/N$. Matching that accuracy requires $d_{b,n,m}^2\le\sigma_k^2/N_{\mathrm{cert}}$. Appendix~\ref{app:reliability} gives conditions for realizing the hidden bias with a valid initial law.
\end{proof}

\paragraph{Implications.}
Even after the diagnostics are satisfied, the squared bias can be at least $d_{b,n,m}^2$. A reported ESS accounts for sampling variance, so it does not remove this lower bound on mean-squared error. Equation~\eqref{eq:certifiable-ess} expresses the accuracy the checklist can guarantee as an equivalent number of independent target draws.

\section{Experiments}
\label{sec:experiments}

We test the residual bound, slow-mode alignment, and anytime guarantees on a synthetic circle and five real-data regression posteriors. Appendix~\ref{app:experiments} gives logistic-posterior comparisons, further synthetic controls, and methods.

\begin{figure*}[t]
\centering
\includegraphics[width=\textwidth]{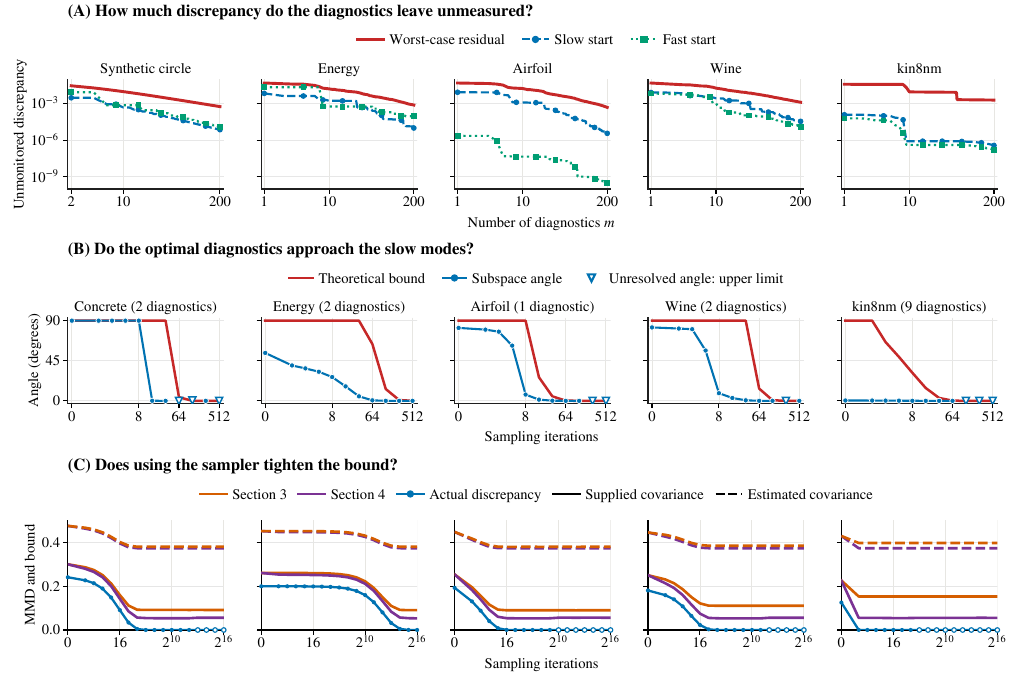}
\vspace{-8pt}
\caption{(A) More diagnostics leave less discrepancy unmeasured. (B) Optimal directions approach the slow modes. (C) Sampler information tightens the bound when covariances are supplied; estimating them limits the gain. Hollow circles mark numerical underflow in the MMD calculation.}
\label{fig:experiments}
\vspace{-6pt}
\end{figure*}

\paragraph{Adding diagnostics.}
\emph{How much discrepancy do additional diagnostics leave unmeasured?} Figure~\ref{fig:experiments}A measures up to 200 leading target-covariance directions while holding each chain's law fixed. The two initializations emphasize slowly and quickly mixing directions. In both cases, the unmonitored discrepancy decreases and stays below the worst-case residual. The synthetic circle has polynomial eigendecay, giving the worst-case residual an asymptotic rate of $m^{-1}$ (Appendix~\ref{app:experiment-residual}).

\paragraph{Choosing diagnostics.}
\emph{Do optimal diagnostics approach the slow modes?} Independent pilots selected $m$ at a gap separating the retained slow modes from the rest, as required by \textbf{(A4)}. The optimal diagnostic span approaches the slow-mode span as sampling proceeds: their angle decreases toward zero, with every numerically resolved angle below the theoretical bound (Figure~\ref{fig:experiments}B). Estimating the slow modes adds the approximation error studied in Section~\ref{sec:approximate-monitors}; longer pilot runs reduce this error (Appendix~\ref{app:experiment-directions}).

\paragraph{Using the anytime bound.}
\emph{Can the bound establish a useful accuracy tolerance?} We compare Sections~3 and~4 on all five regression datasets (Figure~\ref{fig:experiments}C). Both estimate diagnostic readings from independent chains and target draws, using a known initialization radius. Covariances are either supplied or estimated.

Sampler information tightens the bound when covariances are supplied, but covariance uncertainty can erase this gain even after the discrepancy becomes small. The direct MMD bound is nearly as tight as supplied Section~4 at the final inspection and avoids covariance estimation (Table~\ref{tab:regression-bounds}).

\section{Discussion}
\label{sec:discussion}

The anytime bound $\sqrt{M_t^2+U_t^2}$~\eqref{eq:anytime-residual} combines $M_t$, a bound on the monitored discrepancy including estimation uncertainty, and $U_t$, a bound on the unmonitored discrepancy. When estimation uncertainty dominates $M_t$, more observations can help. When $U_t$ dominates, adding diagnostics or choosing better directions can reduce the worst-case residual. Running the sampler longer can reduce the discrepancy itself, while slow modes identify quantities whose bias persists.

Obtaining a useful $U_t$ requires evidence about the discrepancy the checklist does not measure. A justified source class supplies such information, but satisfying the checklist cannot validate that class. Nor can pilots confined to one region reveal all the slow modes involving another. The anytime-bound experiments in Figure~\ref{fig:experiments}C show a further difficulty: uncertainty in the covariance estimates can dominate the bound even when the actual discrepancy is small. Making these bounds informative at an affordable cost remains a practical challenge.

The aim of sampling is usually an accurate posterior estimate. Section~7 shows why its accuracy depends on both bias and variance. Exact convergence need not be certified if these can be bounded tightly enough for the scientific question. Such a guarantee needs more than agreement on a finite checklist: it must also account for the error the checklist cannot see.

\clearpage
\onecolumn
\appendix
\section{Proofs for Section~\ref{sec:certification}}
\label{app:certification}

\subsection{Proof of Theorem~\ref{thm:certification}}
\label{app:qualitative}

We construct a distinct law with the same monitored expectations as $\pi$, then a chain that preserves this agreement at every finite time.

Let $k$ be bounded and characteristic with separable RKHS $\mathcal H$. Write $c_i(x)=\langle v_i,\phi(x)\rangle_{\mathcal H}$ and $c_0\equiv1$. Choose $m+2$ disjoint measurable sets $A_j$ with $\pi(A_j)>0$; these exist when $\pi$ has a density on $\mathbb R^d$. The $(m+1)\times(m+2)$ matrix
\[
M_{ij}=\int_{A_j}c_i(x)\,d\pi(x)
\]
has a nonzero null vector $a$. Scale it so that $\max_j|a_j|=1$ and set $h=\sum_j a_j\mathbf1_{A_j}$. Then $h\ne0$, $|h|\le1$, and $\mathbb E_\pi[c_i(X)h(X)]=0$ for $0\le i\le m$. The constraint for $c_0$ preserves total probability; the others preserve the monitored expectations.

For $0<s<1$, the law $dQ=(1+s h)\,d\pi$ has positive density, integrates to one, and differs from $\pi$. Moreover,
\[
\mathbb E_Q[c_i(X)]-\mathbb E_\pi[c_i(X)]
=s\mathbb E_\pi[c_i(X)h(X)]=0.
\]
Thus $\delta:=\mu_Q-\mu_\pi$ is orthogonal to $V$, and characteristicness gives $\delta\ne0$.

For $0<\rho<1$, use the refresh kernel
\[
K_\rho(x,\cdot)=(1-\rho)\delta_x+\rho\pi,
\]
where $\delta_x$ is point mass at $x$. This kernel is reversible and uniformly ergodic with target $\pi$. Initialized at $Q$, it has law
\[
\pi_t=(1-\rho)^tQ+\bigl[1-(1-\rho)^t\bigr]\pi.
\]
Hence $\delta_t=(1-\rho)^t\delta\ne0$ and $\delta_t\perp V$ at every finite time. The unit witness $v_{m+1}=\delta_t/\|\delta_t\|$ detects $\langle v_{m+1},\delta_t\rangle=\|\delta_t\|=\operatorname{MMD}(\pi_t,\pi)$. Initialized at $\pi$, the same kernel remains stationary and gives the same zero diagnostic readings. Those readings therefore cannot distinguish the two cases.

\subsection{Proofs of Theorems~\ref{thm:quantitative} and~\ref{thm:quantitative-rate}}
\label{app:quantitative}

For rank-one operators, $(a\otimes b)v=a\langle b,v\rangle$. For bounded $k$ and separable $\mathcal H$, $\Sigma$ is positive and trace-class. Its positive-spectrum eigenvectors span $\overline{\operatorname{ran}(\Sigma)}$, which contains $\mathcal B_r(R)$. If $\operatorname{rank}(\Sigma)\le m$, monitoring its range gives width zero. Otherwise, for $\delta=\Sigma^r u$ and $u_j=\langle u,e_j\rangle$,
\[
\|P_{(V_m^\star)^\perp}\delta\|^2
=\sum_{j>m}\lambda_j^{2r}u_j^2
\le R^2\lambda_{m+1}^{2r}.
\]
Equality holds at $\delta=R\lambda_{m+1}^r e_{m+1}$.

For any $V$ with $\dim V\le m$, choose a unit vector $w=\sum_{j=1}^{m+1}w_j e_j$ orthogonal to $V$. Such a vector exists because their $(m+1)$-dimensional span has a nontrivial intersection with $V^\perp$. Set $\delta^\star=R\lambda_{m+1}^r w$. It belongs to $\mathcal B_r(R)$ since
\[
\sum_{j=1}^{m+1}
\frac{|\langle\delta^\star,e_j\rangle|^2}{\lambda_j^{2r}}
=R^2\sum_{j=1}^{m+1}w_j^2
\left(\frac{\lambda_{m+1}}{\lambda_j}\right)^{2r}
\le R^2.
\]
This vector satisfies every diagnostic in $V$ and has norm $R\lambda_{m+1}^r$, proving the matching lower bound. Thus the minimax value is unchanged when the supremum is restricted to $\delta\perp V$. Assumption \textbf{(A2)} immediately gives the rate in Theorem~\ref{thm:quantitative-rate}. These exact widths concern the full source ellipsoid; restricting to discrepancies realized by probability laws can reduce the supremum.

\subsection{Inexact Diagnostics}
\label{app:inexact}

Pythagoras gives~\eqref{eq:residual-decomposition}. For orthonormal diagnostics with $|\langle v_i,\delta_t\rangle|\le\varepsilon$,
\[
\|P_V\delta_t\|^2
=\sum_{i=1}^m|\langle v_i,\delta_t\rangle|^2
\le m\varepsilon^2.
\]
For arbitrary diagnostic directions, let $G_{ij}=\langle v_i,v_j\rangle$, $z_i=\langle v_i,\delta_t\rangle$, and write $G^\dagger$ for the Moore--Penrose pseudoinverse. Then
\[
P_V\delta_t=\sum_i(G^\dagger z)_i v_i,
\qquad
\|P_V\delta_t\|^2=z^\top G^\dagger z.
\]
If $V\ne\{0\}$ and $|z_i|\le\varepsilon$, this gives
\[
\|P_V\delta_t\|
\le\sqrt{\frac{m}{\lambda_{\min}^+(G)}}\,\varepsilon,
\]
where $\lambda_{\min}^+(G)$ is the smallest positive eigenvalue. For $V=\{0\}$ the projection is zero. Thus individual tolerances bound the monitored component through the Gram matrix; orthonormal directions give $G=I$.

When the residual is nonzero, the unit direction
\[
w_t=\frac{P_{V^\perp}\delta_t}{\|P_{V^\perp}\delta_t\|}
\]
is orthogonal to every diagnostic. For $c_{w_t}(x)=\langle w_t,\phi(x)\rangle$, its expectation difference is
\[
\mathbb E_{\pi_t}[c_{w_t}(X)]-\mathbb E_\pi[c_{w_t}(X)]
=\langle w_t,\delta_t\rangle
=\|P_{V^\perp}\delta_t\|.
\]
The residual is therefore an expectation bias attained by an unmonitored RKHS function. In particular, the exact zeros constructed in Appendix~\ref{app:qualitative} also satisfy any positive diagnostic tolerance.

\subsection{When the Source Condition Holds}
\label{app:source-conditions}

Let $\Phi=\phi-\mu_\pi$, $Bh=\mathbb E_\pi[\Phi(X)h(X)]$, and $K=\operatorname*{ess\,sup}_\pi\|\Phi(x)\|$. Then $BB^*=\Sigma$. For $Q\ll\pi$, its centered density ratio $h=dQ/d\pi-1$ gives $\mu_Q-\mu_\pi=Bh$.

\begin{proposition}[Density errors imply source conditions]
\label{prop:lp-source}
If $K<\infty$ and $h\in L^p(\pi)$ for $1<p\le2$, then
\[
\mu_Q-\mu_\pi=\Sigma^{1-1/p}u,
\qquad \|u\|\le K^{2/p-1}\|h\|_p.
\]
Thus \textbf{(A1)} holds with $r=1-1/p$. For $p=2$, the radius is $\|h\|_2=\sqrt{\chi^2(Q\|\pi)}$, the square root of the chi-square divergence \citep{sason2016fdivergence}.
\end{proposition}

\begin{proof}
For $p=2$, polar decomposition gives $B=\Sigma^{1/2}U$ with $\|U\|\le1$. For $1<p<2$, let $P_N$ project onto the first $N$ positive covariance eigendirections and consider $T_z=\Sigma^{-z/2}P_NB$ on $0\le\operatorname{Re}z\le1$. Its boundary norms satisfy
\[
\|T_{iy}\|_{L^1\to\mathcal H}\le K,
\qquad
\|T_{1+iy}\|_{L^2\to\mathcal H}\le1.
\]
Complex interpolation at $\theta=2-2/p$ yields
\[
\|\Sigma^{-(1-1/p)}P_NBh\|
\le K^{2/p-1}\|h\|_p
\]
uniformly in $N$ \citep{stein1956interpolation}. Taking the limit of the squared spectral norms gives the required $u$. No component of $Bh$ lies in $\ker\Sigma$: a centered feature in such a direction vanishes $\pi$-almost surely and hence $Q$-almost surely.
\end{proof}

An $M$-warm law, $dQ/d\pi\le M$, supplies the explicit bound $\chi^2(Q\|\pi)\le M-1$. Indeed, for $q=dQ/d\pi$, $\mathbb E_\pi(q-1)^2=\mathbb E_\pi q^2-1\le M-1$. Multimodality alone need not make the radius large: if $\pi=wQ+(1-w)S$, then $dQ/d\pi\le1/w$. Initializing from one of two equally weighted mixture components therefore gives radius at most one, regardless of component separation. A useful radius still requires a justified density-ratio bound.

A point initialization against a continuous target falls outside \textbf{(A3)}. If a Metropolis sampler rejects at its initial state $x_0$ with probability $q(x_0)>0$, then $\pi_t(\{x_0\})\ge q(x_0)^t>0$ at every finite time. Finite burn-in does not remove this atom. Applying \textbf{(A3)} after burn-in therefore requires a separate argument establishing a finite $L^2(\pi)$ density ratio; randomized initialization is covered when its law satisfies that condition.

\section{Proofs for Section~\ref{sec:spectral-duality}}
\label{app:spectral}

\subsection{Proof of Proposition~\ref{thm:dynamic-width}}
\label{app:dynamic-width}

Let $L^2_0(\pi)$ be the mean-zero subspace of $L^2(\pi)$ and $\Phi=\phi-\mu_\pi$. The embedding operator
\[
B:L^2_0(\pi)\to\mathcal H,\qquad
Bh=\mathbb E_\pi[\Phi(X)h(X)]
\]
has adjoint $B^*v(x)=\langle v,\Phi(x)\rangle$ and satisfies $BB^*=\Sigma$. Boundedness of $k$ makes $B$ Hilbert--Schmidt. The transition operator $\mathcal P$ and its adjoint $\mathcal P^*$ contract $L^2_0(\pi)$.

Density errors evolve as $h_t=(\mathcal P^*)^th_0$. Thus $A_t:=B(\mathcal P^*)^t$ maps initialization errors to discrepancies $\delta_t=A_th_0$, giving the ellipsoid
\[
\mathcal B^{(t)}(R_{\mathrm{init}})
=\{A_th:\|h\|_2\le R_{\mathrm{init}},\ h\in L^2_0(\pi)\}.
\]
The adjoint satisfies $A_t^*v(x)=\langle v,F_t(x)\rangle$, so
\[
A_tA_t^*
=\mathbb E_\pi[F_t(X)\otimes F_t(X)]
=\Sigma_t.
\]
Consequently, the singular values of $A_t$ are $\sqrt{\lambda_{j,t}}$. Applying Appendix~\ref{app:quantitative}'s ellipsoid argument gives width $R_{\mathrm{init}}\sqrt{\lambda_{m+1,t}}$, attained by the leading eigenspace of $\Sigma_t$. The same value holds when discrepancies must satisfy the diagnostics. Missing eigenvalues are zero when the rank is finite. An initial probability law additionally requires $1+h_0\ge0$; upper bounds apply to those laws, while the exact width concerns the full ellipsoid.

At $t=0$, polar decomposition gives $\mathcal B^{(0)}(R_{\mathrm{init}})=\mathcal B_{1/2}(R_{\mathrm{init}})$. Since $A_{t+1}=A_t\mathcal P^*$,
\[
\mathcal B^{(t+1)}(R_{\mathrm{init}})
\subseteq\mathcal B^{(t)}(R_{\mathrm{init}}),
\qquad
\Sigma_{t+1}\preceq\Sigma_t\preceq\Sigma.
\]
This proves the comparison with Section~3's $r=1/2$ bound.

For a reversible sampler, $\mathcal P^*=\mathcal P$ gives $\Sigma_t=B\mathcal P^{2t}B^*$. For $u,v\in\mathcal H$,
\[
\begin{aligned}
\langle u,\Sigma_tv\rangle
&=\langle B^*u,\mathcal P^{2t}B^*v\rangle_{L^2(\pi)}\\
&=\mathbb E_\pi\!\left[
\langle u,\Phi(X_0)\rangle
\langle v,\Phi(X_{2t})\rangle\right].
\end{aligned}
\]
Hence $\Sigma_t$ equals the covariance of stationary features separated by $2t$ transitions. Reversibility is needed for this representation, but not for the width calculation.

\subsection{Proof of Theorem~\ref{thm:slow-optimality}}
\label{app:slow-optimality}

Let $S_m=\operatorname{span}\{\psi_1,\ldots,\psi_m\}$ be the slow transition eigenspace in \textbf{(A4)}, with orthogonal projector $E_m$. Its embedded span is $W_m=BS_m$. The alignment argument parallels \citet[Proposition~4 and Appendix~C.1]{tolkovsky2026mode}. We bound the largest principal angle between $W_m$ and $V_m^\star(t)$, then use Appendix~\ref{app:approximate-monitors} to convert alignment into a relative residual bound.

First, characteristicness makes $B$ injective on $L^2_0(\pi)$. If $Bh=0$ for a nonzero $h$, the positive and negative parts of $h\,d\pi$ have equal mass and equal embeddings. After normalization, characteristicness would make them the same probability law, a contradiction. Since $S_m$ is finite-dimensional,
\[
c_m:=\inf_{\substack{h\in S_m\\\|h\|_2=1}}\|Bh\|>0.
\]
Put $\kappa=\|B\|_{\mathrm{op}}$ and
\[
\gamma_t=\frac{\kappa^2}{c_m^2}\left(\frac ab\right)^{2t},
\]
using $a<b$ from \textbf{(A4)}. By reversibility, we can split
\[
A_t=S+F,\qquad
S=B\mathcal P^tE_m,\quad
F=B\mathcal P^t(I-E_m).
\]
The input subspaces are orthogonal, so $\Sigma_t=SS^*+FF^*$. The range of $S$ is $W_m$, and on this space
\[
SS^*\succeq c_m^2b^{2t}I,\qquad
\|FF^*\|_{\mathrm{op}}\le\kappa^2a^{2t}.
\]
Thus $\lambda_{m,t}\ge c_m^2b^{2t}$ and $\lambda_{m+1,t}\le\kappa^2a^{2t}$. When $\gamma_t<1$, these bounds separate the leading eigenspace from the remainder.

Let $U$ have the leading $m$ orthonormal eigenvectors of $\Sigma_t$ as columns, with eigenvalue matrix $\Lambda$, and put $\Pi=P_{W_m}$. Projecting $\Sigma_tU=U\Lambda$ gives
\[
(I-\Pi)U=(I-\Pi)FF^*U\Lambda^{-1}.
\]
Taking operator norms yields
\begin{equation}
\sin\theta(W_m,V_m^\star(t))\le\gamma_t.
\label{eq:app-dynamic-angle}
\end{equation}
The secant bound in Appendix~\ref{app:approximate-monitors} therefore gives
\begin{equation}
\sup_{\substack{\delta\in\mathcal B^{(t)}(R_{\mathrm{init}})\\
\delta\perp W_m}}\|\delta\|
\le\frac{d_m(t)}{\sqrt{1-\gamma_t^2}}.
\label{eq:app-dynamic-passing}
\end{equation}
Since $\gamma_t\to0$, this proves Theorem~\ref{thm:slow-optimality}. When $d_m(t)>0$, the relative excess is $O((a/b)^{4t})$; if $d_m(t)=0$, the displayed inequality gives zero residual.

\subsection{Approximate Diagnostics and Tolerances}
\label{app:approximate-monitors}

Let $V=V_m^\star$, $d_m=R\lambda_{m+1}^r$, and let $W$ have the same dimension as $V$. The largest principal angle satisfies $\sin\theta=\|P_W-P_V\|_{\mathrm{op}}$ \citep{bjorck1973numerical}.

\begin{proof}[Proof of Corollary~\ref{thm:approximate-monitors}]
For $\delta\perp W$, the principal-angle identity gives $\|P_V\delta\|\le\sin\theta\|\delta\|$. Hence
\[
\cos\theta\|\delta\|
\le\|P_{V^\perp}\delta\|
\le d_m,
\]
where the last inequality is the source-class bound. Dividing by $\cos\theta$ proves the result.
\end{proof}

For $\delta\in\mathcal B_r(R)$ with $\|P_W\delta\|\le\eta$, write $\delta=u+v$ with $u\perp W$ and $v\in W$. Then
\[
\cos\theta\|u\|
\le\|P_{V^\perp}u\|
\le d_m+\sin\theta\|v\|.
\]
Therefore
\begin{equation}
\|P_{W^\perp}\delta\|
\le d_m\sec\theta+\eta\tan\theta.
\label{eq:app-noisy-angle}
\end{equation}
Here $\eta$ bounds the monitored projection norm as in Appendix~\ref{app:inexact}. For the dynamic ellipsoid, replace $d_m$ by $d_m(t)$ and measure the angle to $V_m^\star(t)$. This tolerance term is needed when the approximate directions are used in Section~6.

\subsection{Estimated Covariance Operators}
\label{app:mode-estimation}

Let $\widehat\Sigma_t$ be a positive operator estimate satisfying $\|\widehat\Sigma_t-\Sigma_t\|_{\mathrm{op}}\le\epsilon_\Sigma$, with leading $m$-dimensional space $\widehat W_m$. For a unit vector $v\perp\widehat W_m$,
\[
\langle v,\Sigma_tv\rangle
\le\widehat\lambda_{m+1,t}+\epsilon_\Sigma
\le\lambda_{m+1,t}+2\epsilon_\Sigma.
\]
Since $\Sigma_t=A_tA_t^*$, taking the supremum over such $v$ gives
\begin{equation}
\begin{aligned}
\sup_{\delta\in\mathcal B^{(t)}(R_{\mathrm{init}})}
\|P_{\widehat W_m^\perp}\delta\|
&\le R_{\mathrm{init}}\sqrt{\widehat\lambda_{m+1,t}+\epsilon_\Sigma}\\
&\le\sqrt{d_m(t)^2+2R_{\mathrm{init}}^2\epsilon_\Sigma}.
\end{aligned}
\label{eq:app-estimated-covariance}
\end{equation}
This controls the residual without an eigengap assumption. The first bound also holds whenever $\Sigma_t\preceq\widehat\Sigma_t+\epsilon_\Sigma I$.

Appendix~\ref{app:covariance-estimation} constructs this bound using independent initial states drawn from $\pi$. Estimation from a single trajectory instead requires quantitative dependence assumptions \citep{mollenhauer2022autocovariance}. Agreement between two estimated spaces cannot establish their accuracy if both pilots miss the same region.

\section{Proofs for Section~\ref{sec:no-escape}}
\label{app:trajectory}

\subsection{Shared Discrepancy Across Chains}

Expand $\delta_i=\bar\delta+(\delta_i-\bar\delta)$ in the squared norm and sum over $i$. The cross terms vanish because $\sum_i(\delta_i-\bar\delta)=0$, giving~\eqref{eq:chain-anova}. Pairwise comparisons depend only on $\delta_i-\delta_j$ and cannot detect a component shared by all chains. In particular, identical nonstationary marginal laws give zero disagreement. The same identity applies to laws averaged over different time windows and requires no independence assumption.

\subsection{Nonlinear Functions of Moments}
\label{app:nonlinear-moments}

\begin{proposition}[Nonlinear combinations of finitely many expectations]
\label{prop:nonlinear-moments}
Let $c_1,\ldots,c_q\in L^1(\pi)$, and suppose $\pi$ admits $q+2$ disjoint sets of positive probability. For diagnostics
\[
D_\ell(P)=F_\ell\bigl(\mathbb E_P[c_1],\ldots,\mathbb E_P[c_q]\bigr),
\]
with each $F_\ell$ defined at the target moment vector, there is an ergodic chain targeting $\pi$ such that every $D_\ell(\pi_t)=D_\ell(\pi)$ at every finite time, while $\pi_t\ne\pi$.
\end{proposition}

\begin{proof}
The nullspace construction in Appendix~\ref{app:qualitative}, applied to $1,c_1,\ldots,c_q$, gives a bounded nonzero $h$ satisfying $\mathbb E_\pi h=0$ and $\mathbb E_\pi[h c_j]=0$ for every $j$. Scale it so that $\|h\|_\infty<1$. Then $dQ=(1+h)d\pi$ is a probability law with the same moments as $\pi$. The refresh kernel from that appendix gives $\pi_t-\pi=(1-\rho)^t(Q-\pi)$, so those moments, and hence every diagnostic $D_\ell$, retain their target values at every finite time.
\end{proof}

\paragraph{Taylor (von Mises) expansions.}
Write $dP_h=(1+h)d\pi$. Suppose each map $h\mapsto T_\ell(P_h)$ has an analytic extension to a neighborhood of zero in $L^2_0(\pi)$. Its convergent expansion is
\[
T_\ell(P_h)-T_\ell(\pi)
=\sum_{s=1}^{\infty}\frac1{s!}A_{\ell,s}[h,\ldots,h],
\]
where $A_{\ell,s}$ is the continuous $s$-linear derivative \citep{fernholz1983von}. For an orthonormal basis $\{c_j\}$ of $L^2_0(\pi)$, put $\Delta_j=\langle h,c_j\rangle_2=\mathbb E_{P_h}[c_j]-\mathbb E_\pi[c_j]$. Finite basis projections converge to $h$, giving
\[
\begin{aligned}
A_{\ell,s}[h,\ldots,h]
=\lim_{N\to\infty}\sum_{j_1,\ldots,j_s\le N}
&A_{\ell,s}[c_{j_1},\ldots,c_{j_s}]\\
&{}\cdot\prod_{u=1}^s\Delta_{j_u}.
\end{aligned}
\]
This is the moment expansion used in Section~\ref{sec:nonlinear}. If only finitely many moment functions appear across all orders and diagnostics and the support condition of Proposition~\ref{prop:nonlinear-moments} holds, that proposition applies. Scale its perturbation to remain inside the convergence neighborhood. Every term then vanishes along the refresh chain.

Convergence of the expansion is essential; truncating a nonzero remainder establishes only approximate equality. The number of moment functions also matters, rather than the number of Taylor orders: squared characteristic MMD is second-order but uses infinitely many distribution-determining coordinates.

\paragraph{Infinitely many moments.}
For $c_j\in L^2(\pi)$, let $S=\overline{\operatorname{span}}\{1,c_1,c_2,\ldots\}$. If $S^\perp$ contains a bounded nonzero $h$, scale it to $\|h\|_\infty<1$. The law $dQ=(1+h)d\pi$ then matches every moment while differing from $\pi$. Nondensity supplies an orthogonal direction; boundedness ensures that this construction is a probability law. For example, under the uniform law on the circle, $h(x)=\epsilon\sin x$, $0<|\epsilon|<1$, is orthogonal to the constant and every $\cos(jx)$. Infinitely many cosine criteria therefore leave this sine component undetected.

\subsection{Diagnostics of a Finite Trajectory}

A diagnostic may depend on several states rather than one. Fix a $\pi$-invariant kernel and bounded features $F_1,\ldots,F_p$ of $(X_0,\ldots,X_L)$. Define
\[
g_j(x)=\mathbb E[F_j\mid X_0=x]
-\mathbb E_{X_0\sim\pi}[F_j],
\]
where the second expectation is over a stationary path with the same kernel. For $d\pi_0=(1+h_0)d\pi$, conditioning on the initial state gives
\[
\mathbb E_{X_0\sim\pi_0}[F_j]
-\mathbb E_{X_0\sim\pi}[F_j]
=\langle h_0,g_j\rangle_{L^2(\pi)}.
\]
Thus a finite collection of path expectations imposes finitely many orthogonality constraints on $h_0$.

\begin{proposition}[Finite trajectory diagnostics]
\label{prop:path-features}
If $\pi$ admits $p+2$ disjoint positive-probability sets, there is an initial law $\pi_0\ne\pi$ with bounded positive density relative to $\pi$ that matches all $p$ stationary path expectations. If $\mathcal P^*$ is injective on $L^2_0(\pi)$, then $\pi_t\ne\pi$ at every finite time.
\end{proposition}

\begin{proof}
Apply Appendix~\ref{app:qualitative}'s construction to $1,g_1,\ldots,g_p$. It gives a nonzero $h_0$ with $\|h_0\|_\infty<1$ orthogonal to every constraint. The time-$t$ density error is $(\mathcal P^*)^t h_0$, which remains nonzero under injectivity. The refresh kernel satisfies this condition because $\mathcal P^*=(1-\rho)I$ on mean-zero functions.
\end{proof}

The features here are fixed, and the equalities concern their population expectations. Diagnostics fitted from the same observations require a separate treatment of that fitting procedure.

\section{Proofs for Section~\ref{sec:anytime}}
\label{app:anytime}

\subsection{Proof of Theorem~\ref{thm:anytime}}
\label{app:combined-bound}

On the event where both component bounds hold at every inspection time, Pythagoras gives
\[
\operatorname{MMD}^2(\pi_t,\pi)
=\|P_{V_t}\delta_t\|^2+\|P_{V_t^\perp}\delta_t\|^2
\le M_t^2+U_t^2.
\]
The inequality therefore holds at any inspection time chosen from the data. If the two simultaneous events have failure probabilities $\beta_M$ and $\beta_U$, their intersection has probability at least $1-\beta_M-\beta_U$, without requiring independence.

The baseline and refinement use the residual bounds proved in Appendices~\ref{app:quantitative} and~\ref{app:dynamic-width}. For approximate directions with angle at most $\theta_t<\pi/2$ to the corresponding optimal space, \eqref{eq:app-noisy-angle} gives
\[
U_t=d_m\sec\theta_t+M_t\tan\theta_t,
\]
where $d_m=R\lambda_{m+1}^r$ for Section~3, or $d_m(t)$ for Section~4. Any uncertainty in the angle must be included in the same failure budget.

\subsection{Bounding the Monitored Component}
\label{app:replicas}

Fix $k$ with $k(x,x)\le\kappa^2$. Run $J$ independent chains with the same initialization law and fixed transition kernel. At each predetermined inspection time, form
\[
\widehat\mu_t=\frac1J\sum_{j=1}^J\phi(X_t^{(j)}).
\]
The states are independent across chains at each time; dependence across times is allowed. Choose positive weights $w_t$ with $\sum_t w_t\le1$ over the inspection schedule, and set
\begin{equation}
b_t=\frac{\kappa}{\sqrt J}
\left(1+\sqrt{2\log\frac1{\beta_\mu w_t}}\right).
\label{eq:replica-radius}
\end{equation}
Then, with probability at least $1-\beta_\mu$, $\|\widehat\mu_t-\mu_{\pi_t}\|\le b_t$ at every inspection time.

\begin{proof}
Independence gives $\mathbb E\|\widehat\mu_t-\mu_{\pi_t}\|^2\le\kappa^2/J$, so the expected norm is at most $\kappa/\sqrt J$. Replacing one observation changes that norm by at most $2\kappa/J$. The bounded-differences inequality gives the radius in~\eqref{eq:replica-radius} with failure probability $\beta_\mu w_t$. Sum these probabilities over inspection times.
\end{proof}

Suppose a target reference satisfies $\|\widehat\mu_\pi-\mu_\pi\|\le b_\pi$ with probability at least $1-\beta_\pi$. For $N$ independent target draws, the same argument gives $b_\pi=\kappa(1+\sqrt{2\log(1/\beta_\pi)})/\sqrt N$. On the joint event, let
\[
\widehat\delta_t=\widehat\mu_t-\widehat\mu_\pi,\qquad
\eta_t=b_t+b_\pi.
\]
Projection contraction then gives
\begin{equation}
M_t=\|P_{V_t}\widehat\delta_t\|+\eta_t
\ge\|P_{V_t}\delta_t\|
\label{eq:replica-projected}
\end{equation}
simultaneously for every $t$ and every chosen span $V_t$, including spaces selected from the observations. The total failure probability is at most $\beta_\mu+\beta_\pi$.

Alternatively, for fixed or independently selected orthonormal directions, simultaneous coordinate bounds $|\widehat a_{i,t}-\langle v_{i,t},\delta_t\rangle|\le e_{i,t}$ give
\[
M_t=\left[\sum_i(|\widehat a_{i,t}|+e_{i,t})^2\right]^{1/2}.
\]
These errors must include uncertainty in the corresponding target expectations.

At fixed $J$, running the chains longer does not by itself shrink $b_t$. If replica counts are also selected from the data, allocate error over the inspected time--count pairs. A single-chain average instead estimates the average marginal law over its retained times, the object treated in Section~7.

\subsection{Bounding the Unmonitored Component}
\label{app:covariance-estimation}

Under \textbf{(A3)}, Appendix~\ref{app:mode-estimation} converts a covariance estimate into $U_t$. The following construction estimates $\Sigma_t$ without knowing $\mu_\pi$.

Draw $X,X'\sim\pi$ independently. From each state, run two conditionally independent $t$-step trajectories, with endpoints $Y,Z$ and $Y',Z'$. Set
\[
\begin{aligned}
D_1&=\phi(Y)-\phi(Y'),&
D_2&=\phi(Z)-\phi(Z'),\\
O_t&=\tfrac14(D_1\otimes D_2+D_2\otimes D_1).
\end{aligned}
\]
Conditional independence gives
\[
\mathbb E[O_t\mid X,X']
=\tfrac12(F_t(X)-F_t(X'))^{\otimes2},
\]
so $\mathbb E O_t=\Sigma_t$. This uses stationarity of the target starts and does not require reversibility.

Average $N$ independent copies to obtain $\widehat O_t$. Since $\|O_t\|_{\mathrm{HS}}\le2\kappa^2$, the argument in Appendix~\ref{app:replicas}, applied in the Hilbert space of Hilbert--Schmidt operators, gives
\[
\|\widehat O_t-\Sigma_t\|_{\mathrm{op}}
\le\epsilon_{\Sigma,t}
:=\frac{2\kappa^2}{\sqrt N}
\left(1+\sqrt{2\log\frac1{\beta_\Sigma w_t}}\right)
\]
simultaneously with probability at least $1-\beta_\Sigma$.

Replace negative eigenvalues of $\widehat O_t$ by zero to obtain $\widehat O_{t,+}$. On this event,
\[
\Sigma_t\preceq\widehat O_t+\epsilon_{\Sigma,t}I
\preceq\widehat O_{t,+}+\epsilon_{\Sigma,t}I.
\]
For the leading $m$ eigendirections of $\widehat O_{t,+}$, this one-sided operator bound gives
\[
U_t=R_{\mathrm{init}}
\sqrt{\widehat\lambda_{m+1,t}^{+}+\epsilon_{\Sigma,t}}.
\]
For another diagnostic span $V_t$, replace $\widehat\lambda_{m+1,t}^{+}$ by $\|P_{V_t^\perp}\widehat O_{t,+}P_{V_t^\perp}\|_{\mathrm{op}}$. The event covers all choices of span and diagnostic count.

This construction requires independent initial states drawn from $\pi$, a justified $R_{\mathrm{init}}$, and an error allocation covering every inspected covariance estimate. A finite-feature approximation additionally needs a bound on its omitted component to give a full-RKHS guarantee. The source radius is not estimated from the diagnostic readings.

\subsection{A Bound Without a Source Condition}
\label{app:residual-energy}
\label{sec:measured-residual}

For $\Phi=\phi-\mu_\pi$, define the unmonitored feature energy $g_V(x)=\|P_{V^\perp}\Phi(x)\|^2$. Jensen's inequality gives
\begin{equation}
\|P_{V^\perp}(\mu_Q-\mu_\pi)\|^2
\le\mathbb E_Q[g_V(X)].
\label{eq:residual-energy-bound}
\end{equation}
An upper confidence bound on this additional expectation therefore supplies $U_t^2$ without a source condition.

To include uncertainty in target centering, use a reference $\widehat\mu_\pi$ from Appendix~\ref{app:replicas}, independent of the evaluation trajectories. Fix $V_t$ in advance or select it from an independent pilot, and put
\[
\widetilde g_t(x)
=\|P_{V_t^\perp}(\phi(x)-\widehat\mu_\pi)\|^2.
\]
Conditionally on the reference and pilot, $0\le\widetilde g_t(x)\le L$, where $L=(\kappa+\|\widehat\mu_\pi\|)^2$. Hoeffding's inequality gives the simultaneous bound
\[
G_t=\frac1J\sum_{j=1}^J\widetilde g_t(X_t^{(j)})
+L\sqrt{\frac{\log(1/(\beta_gw_t))}{2J}}
\]
on $\mathbb E_{\pi_t}[\widetilde g_t(X)]$, with failure probability at most $\beta_g$. When the target-reference bound also holds, Jensen's inequality and the triangle inequality give
\[
\|P_{V_t^\perp}\delta_t\|\le\sqrt{G_t}+b_\pi.
\]
All events share Section~6's failure budget.

Even with exact centering, $\mathbb E_\pi g_V(X)=\operatorname{tr}(P_{V^\perp}\Sigma)$ can be nonzero. This bound can therefore remain loose at convergence; it replaces the source condition with an additional measurement and target information.

\section{Proofs for Section~\ref{sec:reliability}}
\label{app:reliability}

\subsection{Proof of Corollary~\ref{cor:reliability}}

For burn-in $b$ and $n$ retained states, define
\[
q_{b,n}(z)=\frac1n\sum_{s=0}^{n-1}z^{b+s},
\qquad
A_{b,n}=Bq_{b,n}(\mathcal P^*).
\]
Averaging the identity $\delta_t=B(\mathcal P^*)^th_0$ gives $\bar\delta_{b,n}=A_{b,n}h_0$. The covariance in Section~7 satisfies
\[
\Sigma_{b,n}=A_{b,n}A_{b,n}^*
=Bq_{b,n}(\mathcal P^*)q_{b,n}(\mathcal P)B^*.
\]
Indeed, $A_{b,n}^*v(x)=\langle v,\bar F_{b,n}(x)\rangle$. The full $L^2_0(\pi)$ ball of radius $R_{\mathrm{init}}$ therefore maps to an ellipsoid with semi-axes $R_{\mathrm{init}}\sqrt{\lambda_{j,b,n}}$. Appendix~\ref{app:quantitative}'s width argument proves~\eqref{eq:average-width}.

For $d_{b,n,m}>0$, fix any $m$-dimensional diagnostic span. The corollary's realizability assumption supplies an initial probability law whose retained-average discrepancy $\bar\delta_{b,n}$ is orthogonal to that span and has norm at least $d_{b,n,m}$. Its unit witness $f=\bar\delta_{b,n}/\|\bar\delta_{b,n}\|$ has bias $\|\bar\delta_{b,n}\|$, so
\[
\operatorname{MSE}(\bar f_{b,n})
\ge\operatorname{Bias}(\bar f_{b,n})^2
\ge d_{b,n,m}^2.
\]
In comparison, the average of $N$ independent target draws is unbiased, with worst-case mean-squared error
\[
\sup_{\|f\|\le1}\frac{\operatorname{Var}_\pi(f(X))}{N}
=\frac{\|\Sigma\|_{\mathrm{op}}}{N}
=\frac{\sigma_k^2}{N}.
\]
A checklist cannot guarantee this accuracy uniformly over the allowed initial laws unless $d_{b,n,m}^2\le\sigma_k^2/N$, proving~\eqref{eq:certifiable-ess}. The initial law and witness may depend on the checklist.

\subsection{Realizing the Lower Bound}

Assume $A_{b,n}$ has rank at least $m+1$, with leading singular values $s_1\ge\cdots\ge s_{m+1}>0$, left singular vectors $u_j\in\mathcal H$, and right singular functions $a_j\in L^2_0(\pi)$. Put
\[
K_{m+1}=\operatorname*{ess\,sup}_{x\sim\pi}
\left(\sum_{j=1}^{m+1}a_j(x)^2\right)^{1/2}.
\]
This constant is finite for bounded $k$: each $a_j=s_j^{-1}q_{b,n}(\mathcal P)B^*u_j$ is bounded because Markov operators contract $L^\infty$.

\begin{proposition}[A sufficient condition for realizability]
If $R_{\mathrm{init}}K_{m+1}\le1$, every diagnostic span of dimension at most $m$ admits an initial probability law satisfying \textbf{(A3)} whose retained-average discrepancy is orthogonal to that span and has norm $d_{b,n,m}$.
\end{proposition}

\begin{proof}
For a diagnostic span $V$, choose a unit vector $w=\sum_{j=1}^{m+1}w_ju_j$ orthogonal to $V$, and set
\[
h=R_{\mathrm{init}}s_{m+1}
\sum_{j=1}^{m+1}\frac{w_j}{s_j}a_j.
\]
Then $\mathbb E_\pi h=0$, $\|h\|_2\le R_{\mathrm{init}}$, and $\|h\|_\infty\le R_{\mathrm{init}}K_{m+1}\le1$. Thus $d\pi_0=(1+h)d\pi$ is a probability law satisfying \textbf{(A3)}. Its retained-average discrepancy is
\[
A_{b,n}h=R_{\mathrm{init}}s_{m+1}w=d_{b,n,m}w,
\]
which satisfies every diagnostic in $V$.
\end{proof}

If the condition fails, the same construction with $R_{\mathrm{real}}=\min\{R_{\mathrm{init}},K_{m+1}^{-1}\}$ still realizes bias $R_{\mathrm{real}}s_{m+1}$. It gives the weaker necessary condition $N_{\mathrm{cert}}\le\sigma_k^2/(R_{\mathrm{real}}^2s_{m+1}^2)$.

\section{Experimental Details}
\label{app:experiments}

Each posterior uses the full dataset, with no training/test split.

\paragraph{Introductory example.}
Figure~\ref{fig:trapped-mixture} uses $\pi=\tfrac12N(8u,I_{10})+\tfrac12N(-8u,I_{10})$, where $u=(1,\ldots,1)/\sqrt{10}$. Four MALA chains start independently from $N(8u,0.01I_{10})$, with proposal standard deviation 0.5 and no discarded warmup. We compute rank-normalized split $\hat R$ and bulk ESS on all ten coordinates, reporting their maximum and minimum.

\paragraph{Synthetic targets.}
The Section~8 experiments used a ten-dimensional Gaussian with covariance condition number 25, an equally weighted Gaussian mixture with component means separated by six standard deviations, a Student-$t$ with three degrees of freedom, and a funnel with $y\sim N(0,3^2)$ and $x\mid y\sim N(0,e^y I_9)$. Gaussian RWM and fixed-covariance MALA used Metropolis--Hastings correction. Coordinate scales were fixed from target median absolute deviations. Independent pilots tuned the step sizes to acceptance rates 0.234 and 0.574, respectively. The kernels were scaled RBF kernels, with bandwidth set to the median distance between independent target draws. Sensitivity tests varied dimension, condition number, mode separation and weights, tail weight, funnel scale, bandwidth, and kernel family.

\paragraph{Gaussian regression posteriors.}
The datasets were Concrete \citep{yeh1998concrete}, kin8nm \citep{openml2014kin8nm}, Airfoil \citep{brooks1989airfoil}, red Wine Quality \citep{cortez2009wine}, and Energy Efficiency \citep{tsanas2012energy}. Energy uses heating load; Wine treats quality as a continuous response. Predictors and responses were standardized using sample standard deviations. An intercept, a $N(0,I)$ coefficient prior, and a fixed noise variance give Gaussian posteriors. We set the noise variance to the least-squares residual variance rounded to two significant figures (Table~\ref{tab:regression-settings}).

Systematic-scan Gibbs preserves Gaussian laws, so we can compute the MMD exactly at every iteration, defined as one update of all coordinates. This sampler is not reversible, but the width and branched covariance construction do not require reversibility (Proposition~\ref{thm:dynamic-width}; Appendix~\ref{app:covariance-estimation}). For posterior covariance $C$, we used
\[
k(x,y)=\exp\!\left[-\frac{1}{2\ell^2}\sum_j\frac{(x_j-y_j)^2}{C_{jj}}\right],
\]
with bandwidth $\ell$ equal to the median scaled distance among 2000 independent posterior draws, fixed before evaluation.

\begin{table}[t]
\centering
\small
\caption{Gaussian-regression settings. Figure~\ref{fig:experiments}C uses the same 18 inspections through 65,536 Gibbs iterations on every dataset. The alignment experiment uses a separate sampler and time grid.}
\label{tab:regression-settings}
\begin{tabular}{lrrrr}
\hline
Dataset & Observations & Predictors & Noise variance & Bandwidth $\ell$\\
\hline
Concrete & 1030 & 8 & 0.39 & 3.6431\\
Energy & 768 & 8 & 0.085 & 3.8980\\
Airfoil & 1503 & 5 & 0.49 & 3.1940\\
Wine & 1599 & 11 & 0.64 & 4.5562\\
kin8nm & 8192 & 8 & 0.59 & 4.0960\\
\hline
\end{tabular}
\end{table}

\paragraph{Logistic regression posteriors.}
The datasets were German Credit \citep{hofmann1994german}, Pima \citep{smith1988pima}, Australian Credit \citep{quinlan1987australian}, Ionosphere \citep{sigillito1989ionosphere}, and Sonar \citep{sejnowski1988sonar}, with $(n,p)=(1000,24)$, $(768,8)$, $(690,14)$, $(351,33)$ and $(208,60)$ observations and predictors, respectively. German Credit uses the numeric encoding; Australian Credit uses the supplied encoding and imputation. We removed Ionosphere's constant second column and retained Pima's recorded zeros without imputation. Predictors were standardized using population standard deviations.

The hierarchical prior was $\alpha\sim N(0,10^2)$, $\sigma\sim\operatorname{HalfCauchy}(1)$, and $\beta_j\mid\sigma\sim N(0,\sigma^2)$. German Credit also used a horseshoe prior with independent half-Cauchy local scales. Positive parameters were log-transformed. A separate calibration run set coordinate scales to $1.4826$ times the median absolute deviation and the RBF bandwidth to the median scaled distance among 2000 draws.

\paragraph{Reference samples.}
For each logistic posterior, 16 NUTS runs \citep{hoffman2014nuts} supplied reference samples, using 5000 warmup steps and 25,000 retained draws per run, target acceptance 0.95, maximum tree depth 10, and a dense mass matrix. Sixteen independent noncentered runs with diagonal mass matrices supplied initial states for the experiments. Starts were uniform on $[-2,2]^d$ in unconstrained coordinates. Additional alignment checks used independent pools of 48 runs for German Credit and 144 for Australian Credit.

The horseshoe reference used noncentering, a diagonal mass matrix, and target acceptance 0.99. A stricter reference used 10,000 warmup steps, target acceptance 0.995, and tree depth 12, retaining 25,000 draws per run. Divergent transitions and disagreement between references limit the horseshoe results. None of the logistic references certifies target expectations; the comparisons depend on their accuracy and the shared calibration data.

\paragraph{Computing infrastructure.}
Experiments ran on a shared node with four NVIDIA RTX 4090 GPUs; analytic calculations and some additional experiments ran on CPUs.

\subsection{Residual versus Diagnostic Count}
\label{app:experiment-residual}

\paragraph{Fixed chain laws.}
Figure~\ref{fig:experiments}A holds each chain's law fixed and adds leading target-covariance diagnostics. We compare the unmonitored discrepancy $\|P_{(V_m^\star)^\perp}\delta_t\|$ with the worst-case residual $R\lambda_{m+1}^r$ for $m=1,\ldots,200$. These laws need not satisfy the added criteria; the bound also controls their projected residual.

On the uniform circle, the kernel has Fourier eigenfunctions $\psi_j$ and paired eigenvalues $\lambda_{2j-1}=\lambda_{2j}=3/(\pi^2j^2)$. For even $m$, the worst-case residual is exactly $R(12/\pi^2)^r(m+2)^{-2r}$. The main figure uses $r=1/2$. We computed the time-ten laws from initial density ratios $1+0.1\operatorname{sign}(\cos x)$ and $1+0.1\operatorname{sign}(\cos 3x)$, both at radius $R=0.1$, under a lazy wrapped-Gaussian sampler with move probability $0.1$ and step standard deviation $0.5$. The slow and fast labels refer to the lowest active frequencies, one and three; they do not order the residuals at every $m$. Periodic quadrature agreed with the computed Fourier tails within $1.6\times10^{-6}$ relative error.

For the regression posteriors, we computed the laws after one Gibbs iteration from Gaussian initial laws with slow and faster mean shifts at the same radius $R_{\mathrm{init}}=0.2$. The bound is $R_{\mathrm{init}}\sqrt{\lambda_{m+1}}$. All 2000 residuals across the five datasets and two initializations lay below it and above numerical uncertainty. Figure~\ref{fig:experiments}A shows Energy, Airfoil, Wine and kin8nm. Spectra were computed using Mehler eigenfunctions with a rank-one centering correction; refining the truncation and quadrature changed the first 201 eigenvalues by at most $1.2\times10^{-13}$ relatively.

\begin{figure*}[t]
\centering
\includegraphics[width=\textwidth]{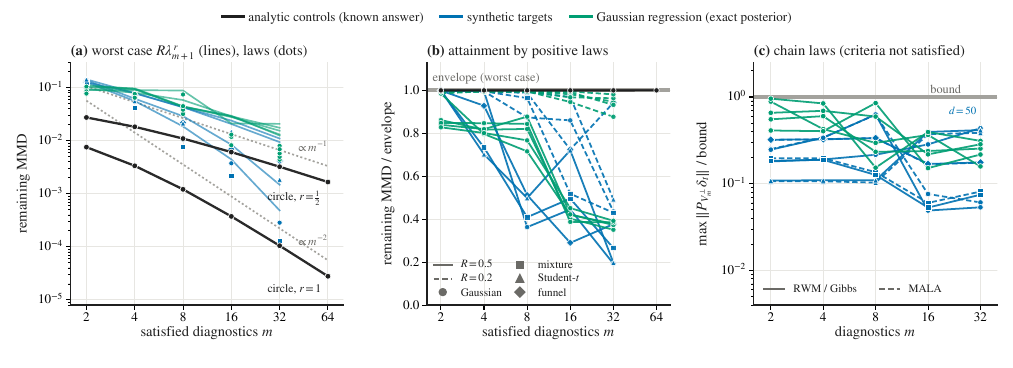}
\caption{Remaining discrepancy as diagnostics are added. (a) Theoretical envelopes (lines) and positive laws satisfying the criteria (dots). (b) Attained fraction of the envelope; positivity and the restricted construction can leave a gap. (c) Projected residuals of actual chain laws, whose criteria need not be satisfied. Regression bounds include numerical allowances. The synthetic ratios in (c) use estimated eigenvalues and are not guarantees.}
\label{fig:r1}
\end{figure*}

\paragraph{Can probability laws attain the bound?}
We separately constructed density ratios $1+h\ge0$ that satisfy the leading $m$ diagnostic criteria, with $\|h\|_{L^2(\pi)}\le R$. Here each value of $m$ uses a different law. Figure~\ref{fig:r1} compares their residuals with the theoretical bound.

On the circle, $1+R\lambda_{m+1}^{r-1/2}\psi_{m+1}$ attains $R\lambda_{m+1}^r$. We evaluated $m=1,\ldots,200$, $r\in\{1/2,1\}$ and $R=0.1$. Each law is realizable at time ten under the same lazy sampler. Independent quadrature with 262,144 points confirmed the residuals to $2.5\times10^{-7}$ relative error.

For two-dimensional versions of the four synthetic families, we made $h$ piecewise constant on $64^2$ equal-probability cells. We imposed normalization and the monitored moment constraints, searched the four largest residual directions with both signs, and rescaled to keep $1+h\ge0.01$. Independent finer quadrature found monitored deviations no larger than $3.7\times10^{-5}$ of the residual. These laws satisfy the criteria numerically, but the search does not certify a maximum. At $R=0.5$, the laws attain at least 99\% of the bound at $m=2$ and 19--38\% at $m=32$. Rescaling for positivity accounts for most of the gap through $m=8$; the finite partition and restricted search also matter at larger $m$.

Regression perturbations used two Gaussian coordinates and integrated out the rest. With $r=1/2$ and $R=0.2$, they attain 88--100\% of the bound for $m\in\{2,4,8,16,32\}$. Figure~\ref{fig:r1}a also shows regression constructions at $R=0.5$. Mat\'ern-3/2 tests on the synthetic families gave similar near-attainment at small radii. For the funnel, this comparison used a scale-mixture surrogate kernel and $m\le16$; the larger diagnostic spans had not numerically converged to the optimal span.

\subsection{Optimal Diagnostics and Slow Modes}
\label{app:experiment-directions}

\paragraph{Known slow modes.}
To test Theorem~\ref{thm:slow-optimality}, we use reversible Gaussian autoregressions whose transition modes and spectral gaps are known. For each regression posterior $N(\mu,C)$, the update is
\[
X'=\mu+A(X-\mu)+\xi,\qquad
\xi\sim N(0,C-ACA^\top),
\]
where $A=C^{1/2}Q\operatorname{diag}(\rho_j)Q^\top C^{-1/2}$ and $Q$ is a fixed random orthogonal matrix. If $s_1\ge\cdots\ge s_d$ are the eigenvalues of the posterior correlation matrix, we set $\rho_j=0.95^{s_1/s_j}$. These samplers have Hermite eigenfunctions and differ from the Gibbs samplers in Figure~\ref{fig:experiments}A and C.

\paragraph{Selecting the diagnostics.}
Four independent stationary pilot chains of length 32,768 selected $m\le16$ at lags $1,4,16,64$. We required a gap larger than five times the half-sample eigenvalue variation \mbox{$\epsilon_m=\frac12\max_{j\le m+2}\bigl||\widehat\rho_j^A|-|\widehat\rho_j^B|\bigr|$}, where $A$ and $B$ denote the two pilot halves. Principal angles had to be below $10^\circ$ between halves and after multiplying the lag by four. We then verified each selected gap against the known transition spectrum. Concrete, Energy and Wine used $m=1,2$, Airfoil $m=1$, and kin8nm $m=9$. Figure~\ref{fig:experiments}B shows $m=2$ where both counts were available and the sole count otherwise.

The calculations use 512 fixed Nystr\"om features, with relative eigenvalue cutoffs $10^{-12}$ for the landmark Gram matrix and $10^{-10}$ for covariance whitening. The population angles decrease toward zero. All numerically resolved angles lay below the theoretical bound, including after adding numerical uncertainty where the sine bound was below one. Angles too small to resolve are shown as hollow markers at their numerical upper limits; a marker above the theoretical curve is not a measured violation. Higher-precision calculations supported the error allowances, but they are not rigorous interval bounds. The comparison is within the 512-feature space.

\begin{figure*}[t]
\centering
\includegraphics[width=\textwidth]{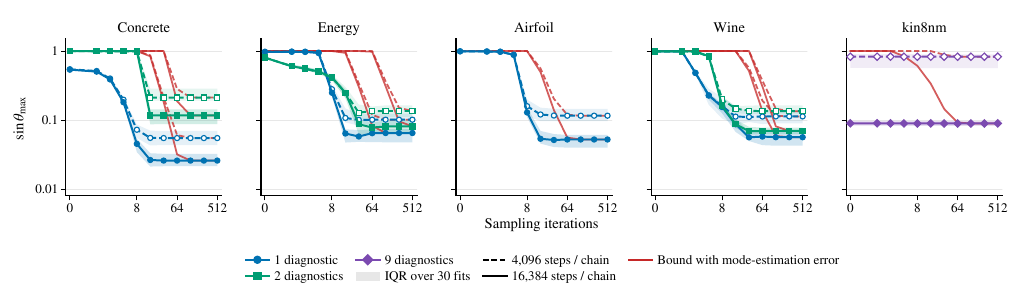}
\caption{Estimating the slow modes leaves an estimation floor. Curves and bands show medians and interquartile ranges over 30 fits. Longer pilots lower this floor. Red curves show median bounds, adding the known error relative to the true slow-mode span to the population envelope; they are not obtainable from the pilot output alone. All quantities use the same 512-feature space as Figure~\ref{fig:experiments}B.}
\label{fig:r2}
\end{figure*}

\paragraph{Estimating the slow modes.}
Figure~\ref{fig:r2} repeats the comparison with modes estimated from 30 sets of four stationary pilot chains. Increasing each chain from 4096 to 16,384 transitions reduced the median angle to the true slow-mode span; on kin8nm it fell from $55.8^\circ$ to $5.1^\circ$. The remaining estimation error makes these curves flatten while the population angles continue to decrease. At $t=64$, satisfying the longer-pilot diagnostics left median worst-case residuals within 0.2\% of the optimum for all eight spans.

For the logistic posteriors, we estimated the optimal spaces from independent reference samples. Longer pilots improved every median angle in Table~\ref{tab:logistic-alignment}. Some comparisons remained numerically unresolved after doubling the reference budget, and we did not verify the transition spectral gap. These results assess empirical alignment, not the theorem's bound. No diagnostic count passed the selection checks for the horseshoe posterior.

\begin{table}[t]
\centering
\small
\caption{Median angles in degrees to reference-estimated optimal spaces on logistic posteriors, using 16,384 transitions per pilot chain. Ranges cover numerically usable inspection times, not uncertainty intervals. All selected diagnostic counts are included; a dash indicates no selected count.}
\label{tab:logistic-alignment}
\vspace{3pt}
\begin{tabular}{lrrr}
\hline
Dataset & $m$ & MALA & RWM\\
\hline
German Credit & 1 & 13.3--15.8 & ---\\
 & 3 & 7.5--8.4 & 26.6--27.3\\
Pima & 1 & 3.9--4.9 & 12.5--13.7\\
 & 3 & 3.3--4.1 & 18.6--20.8\\
Australian Credit & 1 & 5.4--10.4 & 12.6--15.2\\
 & 3 & 11.9--20.1 & 24.2--29.4\\
Ionosphere & 1 & 10.5--11.0 & 19.4--21.0\\
Sonar & 1 & 12.5--13.0 & 28.1--28.4\\
\hline
\end{tabular}
\end{table}

\subsection{Coverage and Usefulness of the Anytime Bound}
\label{app:experiment-anytime}

\begin{table}[t]
\centering
\small
\setlength{\tabcolsep}{3pt}
\caption{Median bounds at 65,536 Gibbs iterations, from 500 repetitions per dataset in Figure~\ref{fig:experiments}C. Supplying the covariance reveals Section~4's improvement; estimation uncertainty largely obscures it. The direct MMD bound uses the same replicas and target reference without estimating a covariance.}
\label{tab:regression-bounds}
\vspace{3pt}
\begin{tabular}{lrrrrr}
\hline
& \multicolumn{2}{c}{Supplied} & \multicolumn{2}{c}{Estimated} & Direct\\
Dataset & Sec.~3 & Sec.~4 & Sec.~3 & Sec.~4 & MMD\\
\hline
Concrete & .092 & .056 & .380 & .373 & .057\\
Energy & .090 & .053 & .380 & .373 & .057\\
Airfoil & .090 & .056 & .380 & .373 & .057\\
Wine & .111 & .056 & .385 & .373 & .057\\
kin8nm & .153 & .056 & .397 & .373 & .057\\
\hline
\end{tabular}
\end{table}

\begin{figure*}[t]
\centering
\includegraphics[width=\textwidth]{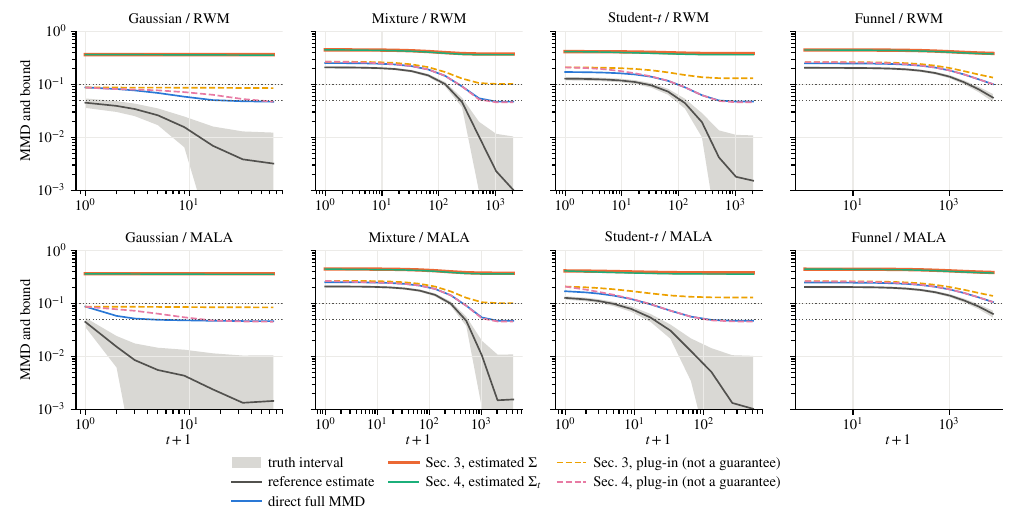}
\caption{Anytime bounds on the four synthetic families under RWM and MALA, with $R_{\mathrm{init}}=0.8$, $J=32768$, $m=8$, and 1000 repetitions. Operator uncertainty prevents the estimated-operator bounds from reaching 0.10. The direct bound remains informative; dashed plug-in curves omit operator uncertainty and are not guarantees. The MMD is shown with its separate reference interval.}
\label{fig:r3-synthetic}
\end{figure*}

\paragraph{Comparison.}
We compare Section~3's static bound with two versions of Section~4's bound: one uses the same diagnostics, and the other uses its own optimal span. Two further comparisons show what each source of information contributes: the direct bound $\|\widehat\delta_t\|+\eta_t$ uses observations without a covariance estimate, while $R_{\mathrm{init}}\sqrt{\lambda_{1,t}}$ uses only the initialization class and sampler. Covariances are either supplied analytically or estimated from independent branched simulations (Appendix~\ref{app:covariance-estimation}). Both variants estimate the diagnostic readings. We also evaluate plug-in formulas that omit covariance uncertainty; these are not confidence bounds.

\paragraph{Sampling and confidence levels.}
Each repetition used fresh chain replicas, target draws, and covariance simulations. We allocated failure probability 0.05 across all inspections and replica counts within a setting. For the synthetic targets, the allocations to replicas, target expectations and covariance estimates were 0.02, 0.01 and 0.02; for regressions, they were 0.025, 0.0125 and 0.0125. The Hilbert-norm bound covers all diagnostic spans on the same event. Coverage is simultaneous over inspections, not across experiment settings.

Figure~\ref{fig:experiments}C shows pointwise median bounds from 500 repetitions per regression dataset. Each repetition uses $J=16384$ independent chain replicas, $m=8$ diagnostics, 2048 covariance blocks and 65,536 independent target draws. All five datasets share the 18 inspection times $0,1,2,4,\ldots,65536$; the replica and covariance failure budgets are divided equally over these times. We simulate the exact Gaussian transition between inspections, preserving their joint law without simulating every intervening Gibbs iteration.

The initial mean shift follows a slow direction with known radius $R_{\mathrm{init}}=0.8$, computed from \mbox{$R_{\mathrm{init}}^2=\exp(a^\top C^{-1}a)-1$} for shift $a$ and posterior covariance $C$. Supplied covariances use 2048 landmarks, a relative Gram cutoff of $10^{-8}$ and a numerical allowance of $10^{-8}$. Additional comparisons across replica budgets and initializations used 1000 repetitions on Concrete and 500 on each other dataset, with $J\in\{1024,4096,16384\}$, $m\in\{2,8,32\}$, and slow and faster shifts at the same radius. Those comparisons used 64, 4096, 16, 32 and 1 iterations on Concrete, Energy, Airfoil, Wine and kin8nm, respectively.

For each of the four synthetic targets and two samplers, 1000 repetitions used $2^{16}$ target draws, nested replica counts $J\in\{4096,8192,16384,32768\}$, $N=2048$ covariance blocks, and $m\in\{2,8,32\}$. The initial density ratio was $1+0.8(2\mathbf1_A-1)$ with $\pi(A)=1/2$, giving radius 0.8. We chose $A$ as a half-space for the Gaussian and mixture, the central radial half for Student-$t$, and $\{y>0\}$ for the funnel. Inspections were at zero and powers of two up to an independently selected horizon, capped at 8192. The circle used 2000 repetitions.

\paragraph{Assessing coverage.}
The regression chain laws give exact MMD values; the circle uses exact enclosures. For the other synthetic targets, an independent run with $2^{18}$ replicas gave reference intervals $[L_t^{\mathrm{ref}},U_t^{\mathrm{ref}}]$ with failure probability 0.01. We counted a bound as covered only when it exceeded $U_t^{\mathrm{ref}}$ at every inspection. Comparisons falling inside the interval remained unresolved. Coverage for these targets is conditional on the reference intervals being valid.

Numerical error also matters. Regression bounds include omitted-trace bounds and numerical allowances. At late inspections, direct evaluation of the regression MMD squared can underflow; we retain the log-scale value and mark these cases with hollow circles in Figure~\ref{fig:experiments}C. They do not indicate exact convergence. Student-$t$ and funnel calculations use a quadrature tolerance of $10^{-10}$, included in both reference intervals and bounds using the known embedding. This tolerance was checked independently but not certified. Eigensolver enclosures and Gram corrections were included; Gram-matrix formation error was certified for some calculations and checked numerically for the rest. The reported coverage depends on these numerical allowances.

\paragraph{What limits useful stopping?}
Every evaluated bound covered the exact MMD at all 18 inspections in 500/500 repetitions per regression dataset. With supplied covariances, Section~4 reached 0.10 in every repetition on all five datasets; Section~3 reached it only on Concrete, Energy and Airfoil. Section~4 first crossed at iterations 32, 8192, 4, 16 and one on Concrete, Energy, Airfoil, Wine and kin8nm, respectively. Section~3 crossed one inspection later on each of the three datasets where it reached the tolerance.

Neither estimated-covariance method reached 0.10. At $R_{\mathrm{init}}=0.8$ and $N=2048$, covariance uncertainty alone imposes the floor
\[
U_t\ge R_{\mathrm{init}}\sqrt{\epsilon_{\Sigma,t}}\approx0.37.
\]
For this confidence construction, the term decays as $N^{-1/4}$: at the same failure allocation, it would take roughly $3.8\times10^5$ blocks just to bring it below 0.10. This is not a lower bound for other estimators. The direct bound avoids this covariance estimate and reached 0.10 in every regression repetition. On the synthetic targets at $J=32768$, it reached 0.10 in all 1000 repetitions on the Gaussian, mixture and Student-$t$ targets. On the funnel, 323/1000 RWM and 3/1000 MALA runs reached it by the horizon. The Gaussian's initial MMD was already below 0.10.

The bound using only the initialization class and sampler reached 0.10 at least as early as Section~4's observed bound for these slow shifts. Observations helped more in the additional comparisons with faster shifts: on Concrete they brought Section~4's first crossing from 32 iterations to one. A post hoc circle initialization proportional to $\exp(\cos3x)$, with $m=8$ and $J=2^{16}$, reduced the crossing from 128 steps to eight.

\paragraph{Computational cost.}
Section~4 estimates $\Sigma_t$ at each inspection; Section~3 can reuse $\Sigma$. We charged for transitions, kernel evaluations and eigendecompositions, giving Section~3 the largest tested replica budget that cost no more than Section~4. At small budgets, Section~3 with more replicas was as tight or tighter on the Gaussian and funnel, and within 0.0004 on the mixture. On Student-$t$, Section~4 fell below Section~3's unmonitored floor, so more replicas alone could not match it at the same covariance estimate and diagnostic count. Costs were modeled from shared-GPU timings, not measured on exclusive hardware.

With $q$ fixed features, accumulating $N$ covariance blocks costs $O(Nq^2)$ time and $O(q^2)$ working memory when processed sequentially. Dense eigendecomposition costs $O(q^3)$ time and $O(q^2)$ memory; projecting $J$ replicas onto $m$ diagnostics costs $O(Jqm)$ once their features are computed. Stepwise branched estimation at time $t$ also requires $O(Nt)$ transitions; the exact Gaussian transitions avoid this cost in the common-horizon experiment. These counts exclude feature construction and the cost per transition.

\paragraph{Sensitivity to the initialization radius.}
Across 60 sensitivity settings with 200 repetitions each, the confidence bounds retained coverage at every inspection. At radius 0.8, estimated-covariance bounds still never reached 0.10. Reducing the radius to 0.2 allowed Section~4 to reach it in 191/200 Gaussian MALA runs, 134/200 Gaussian RWM runs, and 4/200 Student-$t$ MALA runs at $J=32768$, $m=8$, $N=2048$. Section~4's mean final bound was smaller in every setting, although this ordering did not hold at every earlier inspection.

We also tested an incorrect radius of 0.8 in 16 settings, with 100 repetitions each. Overdispersed starts used $\mu+3(X-\mu)$ for $X\sim\pi$; point starts lay three fixed coordinate scales from $\mu$ along the widest coordinate. The assumed radius was invalid for every initial law. Radius-dependent formulas then fell below the discrepancy on the Gaussian, mixture and Student-$t$ targets, while the direct bound retained coverage. These failures do not violate Theorem~\ref{thm:anytime}: its initialization assumption was false. For overdispersed Student-$t$ starts, using the actual finite radius $R_{\mathrm{init}}=2.819$ restored coverage at all inspections in 100/100 repetitions under both samplers.

\paragraph{Using fewer chains.}
For the five hierarchical logistic posteriors, we used RWM and MALA with overdispersed and box initializations. With 14--16 inspections, the replica-error term alone was about 2.44, 1.22 and 0.61 for 4, 16 and 64 chains. Under this error allocation, over 2300 replicas are needed to bring that term below 0.10, before target-reference or covariance uncertainty. This sample requirement belongs to the confidence construction, not to MCMC generally.

Omitting the uncertainty allowances makes stopping easier but can give misleading results. We evaluated the Section~4 plug-in formula with eight estimated directions and an assumed radius of one. Diagnostic estimation and held-out evaluation used disjoint replicas and separate NUTS references. Among 640 groups of 64 chains, 637 crossed 0.10; the held-out discrepancy exceeded 0.10 at 198 crossings and was unresolved at another 136 (Table~\ref{tab:logistic-stopping}). Both the retained-feature discrepancy and full MMD gave these counts. The comparison uses pointwise jackknife uncertainty, not an anytime guarantee, and depends on uncertified references and radii. Independent replication on German Credit, Pima and Australian Credit found 58 exceedances among 240 crossings, with 66 unresolved.

\begin{table}[t]
\centering
\small
\setlength{\tabcolsep}{4pt}
\caption{Plug-in stopping on logistic posteriors. Each dataset has 128 groups across both samplers and initializations. At the first crossing of 0.10, ``above'' means the held-out estimate of MMD$^2$ minus two standard errors exceeds $0.10^2$. Three Ionosphere groups never cross. These are reference-based checks, not failures of a valid confidence bound.}
\label{tab:logistic-stopping}
\vspace{3pt}
\begin{tabular}{lrrr}
\hline
Dataset & Crossings & Above & Unresolved\\
\hline
German Credit & 128 & 49 & 30\\
Pima & 128 & 12 & 46\\
Australian Credit & 128 & 45 & 30\\
Ionosphere & 125 & 41 & 17\\
Sonar & 128 & 51 & 13\\
\hline
Total & 637 & 198 & 136\\
\hline
\end{tabular}
\end{table}

\paragraph{Data and software licenses.}
The cited UCI records for Concrete, Energy, Airfoil, Wine, German Credit, Australian Credit, Ionosphere and Sonar specify \href{https://creativecommons.org/licenses/by/4.0/}{CC BY 4.0}. The OpenML records for \href{https://www.openml.org/api/v1/json/data/189}{kin8nm} and \href{https://www.openml.org/api/v1/json/data/37}{Pima} label their licenses ``Public'' without naming a specific license. Computation used \href{https://github.com/numpy/numpy/blob/main/LICENSE.txt}{NumPy}, \href{https://github.com/scipy/scipy/blob/main/LICENSE.txt}{SciPy}, \href{https://github.com/pytorch/pytorch/blob/main/LICENSE}{PyTorch}, \href{https://github.com/pandas-dev/pandas/blob/main/LICENSE}{pandas}, \href{https://github.com/mpmath/mpmath/blob/master/LICENSE}{mpmath} and \href{https://github.com/omry/omegaconf/blob/master/LICENSE}{OmegaConf} (BSD licenses); \href{https://github.com/jax-ml/jax/blob/main/LICENSE}{JAX}, \href{https://github.com/pyro-ppl/numpyro/blob/master/LICENSE.md}{NumPyro} and \href{https://github.com/arviz-devs/arviz/blob/main/LICENSE}{ArviZ} (Apache 2.0); and \href{https://github.com/facebookresearch/hydra/blob/main/LICENSE}{Hydra} and \href{https://github.com/yaml/pyyaml/blob/main/LICENSE}{PyYAML} (MIT). Figures used \href{https://matplotlib.org/stable/project/license.html}{Matplotlib under its own license}.

\paragraph{Data use.}
The credit and diabetes benchmarks contain financial and health attributes. The modeled arrays contain no names or direct identifiers. We use these datasets only to construct posterior distributions for evaluating samplers, not to make decisions about individuals. We did not collect new participant data or verify the original participant-consent procedures.

\end{document}